\documentclass[twocolumn,showpacs,preprintnumbers,pra]{revtex4-1}
\usepackage{amssymb}
\usepackage{amsmath}
\usepackage{graphicx}
\usepackage{dcolumn}
\usepackage{bm}

\begin{document}

\title{Quantum Coherence via Skew Information and Its Polygamy}
\author{Chang-shui Yu}
\email{ycs@dlut.edu.cn;quaninformation@163.com}
\affiliation{School of Physics and Optoelectronic Technology, Dalian University of
Technology, Dalian 116024, China }
\date{\today }

\begin{abstract}
Quantifying coherence is a key task in both quantum mechanical theory and
practical applications. Here, a reliable quantum coherence measure is
presented by utilizing the quantum skew information of the state of interest
subject to a certain broken observable. This coherence measure is proven to
fulfill all the criteria (especially the strong monotonicity) recently
introduced in the resource theories of quantum coherence. The coherence
measure has an analytic expression and an obvious operational meaning
related to quantum metrology. In terms of this coherence measure, the
distribution of the quantum coherence, i.e., how the quantum coherence is
distributed among the multiple parties is studied and a corresponding
polygamy relation is proposed. As a further application, it is found that
the coherence measure forms the natural upper bounds for quantum
correlations prepared by incoherent operations. The experimental measurement of our coherence measure as well as
the relative-entropy coherence, $l_p$-norm coherence is studied finally.
\end{abstract}

\maketitle

\section{Introduction}

Quantum coherence stemmed from the state superposition principle is the most
fundamental feature of quantum mechanics that distinguishes the quantum from
the classical world. It is the root of all the other intriguing quantum
features such as entanglement \cite{horoe}, quantum correlation \cite%
{disc1,disc2}, quantum non-locality and so on \cite{bell}. Coherence is also
a vital physical resource with various applications in biology \cite%
{Engel,Plenio,Coll,loyd,licm,Huel}, thermodynamical systems \cite%
{Ryb,berg,Nar,Horo,Los1,Los2}, transport theory \cite{reb,wit} and nanoscale
physics \cite{Vaz,Karl}. Since the seminal work \cite{Pleniom} defined the
ingredients in the quantification of coherence such as the "incoherent
states", the "incoherent operations" and the criteria (null, monotonicity
and convexity) of a good coherence measure for the resource theory, quantum 
coherence has attracted increasing interest in many aspects ranging from the
coherence measures \cite{Pleniom, Giro, Napoli, Lewen}, the different
understandings of coherence \cite{Yu09,Stre,Ma,yuto}, and especially the
operational resource theory \cite{Winter,Chi,Chi2,Chi3} and so on \cite%
{Marvian,Marvian2,Yao,Sing,Rast,Piani,Radha} (and references therein).

However, the coherence research is still quite limited. Coherence measure,
first as a mathematical quantifier, has been only well understood based on
the relative entropy and $l_1$ norm especially considering the strong
monotonicity and the closed expression, while for experimental practice,
only the relative-entropy coherence can be, in principle, exactly measured
without the full quantum state tomography (QST) \cite{tom1,tom2} (shown in
Appendix \ref{app:measure}), even though the measurable bounds can be found for other
coherence quantifiers such as the measure based on the $l_1$ norm (given in
this paper) and the robustness of coherence (ROC) \cite{Napoli}. In fact,
different quantifications of coherence can greatly enrich our understanding
of coherence. For example, the relative-entropy coherence can be understood
as the optimal rate for distilling a maximally coherent state from given
states \cite{Winter}. ROC is shown to quantify the advantage enabled by a
quantum state in a phase discrimination task \cite{Napoli}. But the attempt
based on quantum skew information (QSI) failed to quantify the coherence of
a general state \cite{Giro} (shown in the Appendix \ref{app:mono}, also found in Ref. \cite%
{Marvian2}), even though the Wigner-Yanase skew information \cite%
{skew1,skew2,skewl}  and the quantum Fisher information \cite%
{fisher1,fisher2} are more accessible measures of relevance for quantum
metrology as mentioned in \cite{Napoli}. So besides the expected operational
meaning, how to revive the skew information for coherence measure is also of
vital mathematical significance.

In addition, the relative-entropy coherence measure has been shown to be
closely related to the entanglement \cite{Stre} which has an important
characteristic------the monogamy, that is, the entanglement in a
multipartite system can not be freely shared by several subsystems \cite%
{mono1,mono2,mono3} (and references therein). The simplest example is that
once three qubits are maximally entangled, any two qubits among them cannot
own any entanglement, or equivalently, two maximally entangled qubits are
prohibited from entangling with the third qubit. Similarly, is the coherence
freely shared among multipartite system? Recently, the relative-entropy
coherence with free reference basis was studied for multipartite systems in
Ref. \cite{Yao, Radha}, in particular, Ref. \cite{Radha} constructed the
tradeoff relation (monogamy or polygamy) not only depending on the state but
also accompanied by the basis-free coherence. How is the coherence
distributed in terms of a different measure, especially completely by the
basis-dependent measure (as the original purpose of coherence measure)? It
is of immense importance to solve this question for understanding coherence
both as a quantum mechanical feature and as a useful physical resource.

In this paper, we employ quantum skew information to construct a novel
quantum coherence measure which is valid for any quantum state. The most
prominent advantage is that this coherence measure satisfies the strong
monotonicity. Another advantage is that the coherence has an analytic
(closed) expression which is similar to the relative-entropy coherence and $l_1$-norm coherence, but different from the non-analytic ROC \cite{Napoli}.  We
employ this coherence measure to construct a clear polygamy relation that
dominates the coherence distribution among multipartite systems. As a
further application, we consider the tradeoff relation between quantum
coherence and quantum discord and find the natural upper bounds of quantum
discord.  Furthermore, our coherence measure inherits the property of QSI,
so a close relation with the quantum metrology is founded. Finally the
measurement for the experimental practice is considered for various coherence measures.

\section{Coherence via QSI}

To begin with, we would like to first introduce the strict definition of
coherence \cite{Pleniom}. Given a reference basis $\left\{ \left\vert
i\right\rangle \right\} $, a state $\hat{\delta}$ is incoherent if $\hat{%
\delta}=\sum\limits_{i}\delta _{i}\left\vert i\right\rangle \left\langle
i\right\vert $. The states with other forms are coherent. The
incoherent state set is denoted by $\mathcal{I} $. The incoherent operations
are defined by the incoherent completely positive and trace preserving
mapping (ICPTP), i.e., the Kraus operator $\sum_{n}K_{n}^{\dag }K_{n}=%
\mathbb{I}$, if $K_{n}\sigma _{I}K_{n}^{\dag }\in \mathcal{I}$ for $\forall
\sigma _{I}\in \mathcal{I}$. Thus a good coherence measure $C\left( \rho
\right) $ of the state $\rho $ should 

(a) (null) be zero for incoherent
states;

 (b1) (strong monotonicity) not increase under selective ICPTP $%
\$_{I}\left( \rho \right) =\sum_{n} K_{n}\rho K_{n}^{\dag }$ and i.e., $%
C\left( \rho \right) \geq \sum_n p_nC\left( \rho_n\right) $ with $p_{n}=%
\mathrm{Tr}K_{n}\rho K_{n}^{\dag }$ and $\rho_n=K_{n}\rho K_{n}^{\dag }/p_n$%
; 

(b2) (monotonicity) not increase under ICPTP, i.e., $C\left( \rho \right)
\geq C\left( \$_{I}\left( \rho \right) \right) $; 

(c) (convexity) not
increase under classically mixing, i.e., $\sum\limits_{n}q_{n}C\left(
\varrho _{n}\right) \geq C\left( \varrho \right) $ with $\varrho
=\sum_{n}q_{n}\varrho _{n}$, $\sum_{n}q_{n}=1$, $q_{n}>0$. 

It is obvious that in such a framework the definition of coherence strongly depends on the basis. This can be easily understood because the bases could not be arbitrarily changed in the practical scenario. For example, in an experiment the standard Control-Not (CNOT) gate of two qubits  takes the right effect only within some fixed bases. Thus the CNOT gate can transform the coherent joint state $\frac{1}{\sqrt{2}}(\left\vert 0\right\rangle+\left\vert 1\right\rangle)\left\vert 0\right\rangle$ to the maximally entangled state $\frac{1}{\sqrt{2}}\left(\left\vert 00\right\rangle+\left\vert 11\right\rangle\right)$, but do nothing on the incoherent joint state $\left\vert 0\right\rangle\left\vert 0\right\rangle$ \cite{Stre}. This provides an explicit meaning for the basis dependence of the coherence. 

Since the states without off-diagonal entries in the basis are incoherent,
the usual and intuitive way to quantifying the coherence is to measure the
distance between the given state and its closest incoherent state according
to different (pseudo-) distance norms, as done in almost all the mentioned
coherence measures above. In fact, whether the density matrix is diagonal or not in a basis can be directly revealed by the commutation relation between the density matrix of interest and the given (non-degenerate) observable which equivalently (unambiguously) determines a
group of basis. In the following, we establish our coherence measure just by quantifying to what degree the density matrix doesn't commute with some given (broken) observable.

\textbf{Theorem 1}.-The quantum coherence of $\rho $ in the computational
basis $\left\{ \left\vert k\right\rangle \right\} $ can be quantified by%
\begin{equation}
C\left( \rho \right) =\sum\limits_{k=0}^{N_{D}-1}I\left( \rho ,\left\vert
k\right\rangle \left\langle k\right\vert \right) ,  \label{defi}
\end{equation}%
where $I\left( \rho ,\left\vert k\right\rangle \left\langle k\right\vert
\right) =-\frac{1}{2}\mathrm{Tr}\left\{ \left[ \sqrt{\rho },\left\vert
k\right\rangle \left\langle k\right\vert \right] \right\} ^{2}$ represents
the skew information subject to the projector $\left\vert k\right\rangle
\left\langle k\right\vert $ ($N_{D}-1$ is usually omitted if no confusion
occurs). $C\left( \rho \right) $ is a strongly monotonic coherence measure.

Before the proof of the theorem 1, we first introduce two very useful lemmas.

\textbf{Lemma 1}.- Define the function $f(\rho ,\sigma )=\mathrm{Tr}\sqrt{%
\rho }\sqrt{\sigma }$ for arbitrary two density matrices $\rho $ and $\sigma 
$, and the coherence $C\left( \rho \right) $ can be expressed as 
\begin{eqnarray}
C\left( \rho \right) &=&1-\sum_k\left\langle k\right\vert\sqrt{\rho}%
\left\vert k\right\rangle^2  \label{diyib} \\
& =&1-\left[ \max_{\hat{\delta}\in \mathcal{I}}f\left( \rho ,\hat{\delta}%
\right) \right] ^{2}.  \label{bianxing}
\end{eqnarray}%
In particular, $\hat{\delta}=\hat{\delta}^{o}=\sum_{k}\frac{\left\langle
k\right\vert \sqrt{\rho }\left\vert k\right\rangle ^{2}}{\sum\limits_{k^{%
\prime }}\left\langle k^{\prime }\right\vert \sqrt{\rho }\left\vert
k^{\prime }\right\rangle ^{2}}\left\vert k\right\rangle \left\langle
k\right\vert $ is the optimal incoherent state that achieves the maximal
value.

\textbf{Proof}. At first, one can easily find that Eq. (\ref{diyib}) is
valid by expanding $I\left( \rho ,\left\vert k\right\rangle \left\langle
k\right\vert \right) $ in Eq. (\ref{defi}). So the details are omited here.

Next, let's prove Eq. (\ref{bianxing}). Within the computational basis $%
\left\{ \left\vert k\right\rangle \right\} $, the incoherent state $\hat{%
\delta}$ can be explicitly written as 
\begin{equation}
\hat{\delta}=\sum_{k=0}^{N_{D}-1}\hat{\delta}_{kk}\left\vert k\right\rangle
\left\langle k\right\vert .  \label{inbian}
\end{equation}%
Thus we have 
\begin{eqnarray}
f\left( \rho ,\hat{\delta}\right)  &=&\sum_{k=0}^{N_{D}-1}\left\langle
k\right\vert \sqrt{\rho }\left\vert k\right\rangle \sqrt{\hat{\delta}_{kk}} 
\notag \\
&=&Q\sum_{k=0}^{N_D-1}\frac{\left\langle k\right\vert \sqrt{\rho }\left\vert
k\right\rangle }{Q}\sqrt{\hat{\delta}_{kk}}  \label{ree1}
\end{eqnarray}%
with $Q=\sqrt{\sum_{k=0}^{N_{D}-1}\left\langle k\right\vert \sqrt{\rho }%
\left\vert k\right\rangle ^{2}}$. According to the Cauchy-Schwarz
inequality, we have 
\begin{eqnarray}
&&\left( \sum_{k=0}^{N_{D}-1}\frac{\left\langle k\right\vert \sqrt{\rho }%
\left\vert k\right\rangle }{Q}\sqrt{\hat{\delta}_{kk}}\right) ^{2}  \notag \\
&\leq &\left( \sum_{k=0}^{N_{D}-1}\frac{\left\langle k\right\vert \sqrt{\rho 
}\left\vert k\right\rangle ^{2}}{Q^{2}}\right) \left(
\sum_{k=0}^{N_{D}-1}\hat{\delta}_{kk}\right) =1  \label{ineqs}
\end{eqnarray}%
with the inequality saturated for 
\begin{equation}
\sqrt{\hat{\delta}_{kk}}=\frac{\left\langle k\right\vert \sqrt{\rho }%
\left\vert k\right\rangle }{Q}.  \label{tiaojian}
\end{equation}%
Substitute Eq. (\ref{ineqs}) into Eq. (\ref{ree1}), one will find 
\begin{equation*}
f\left( \rho ,\hat{\delta}\right) \leq Q,
\end{equation*}%
or 
\begin{equation}
\left[ \max_{\hat{\delta}\in \mathcal{I}}f\left( \rho ,\hat{\delta}\right) %
\right] ^{2}=Q^{2}=\sum_{k=0}^{N_{D}-1}\left\langle k\right\vert \sqrt{\rho }%
\left\vert k\right\rangle ^{2}.  \label{ree3}
\end{equation}%
Comparing Eq. (\ref{diyib}) and Eq. (\ref{ree3}), one can immediately find
that our Eq. (\ref{bianxing}) is satisfied.

In addition, since Eq. (\ref{tiaojian}) saturates Eq. (\ref{ineqs}), one can
find the optimal incoherent state can be directly obtained by substituting
Eq. (\ref{tiaojian}) into Eq. (\ref{inbian}), which completes the proof.
\hfill$\blacksquare$

\textbf{Lemma 2.}-Let $\$=\{M_{n}\}$ denote any quantum channel given in the
Kraus representation with $\sum\limits_{n=0}M_{n}^{\dag }M_{n}=\mathbb{I}$,
then for any two density matrices $\rho$ and $\sigma$, 
\begin{equation}
f(\rho ,\sigma )\leq\sum_{n}\sqrt{p_{n}q_{n}}f\left( \rho _{n},\sigma
_{n}\right) ,  \label{lem2}
\end{equation}%
with $p_{n}=$Tr$M_{n}\rho M_{n}^{\dag }$, $q_{n}=$Tr$M_{n}\sigma M_{n}^{\dag
}$ and $\rho _{n}=M_{n}\rho M_{n}^{\dag }/p_{n},\sigma _{n}=M_{n}\sigma
M_{n}^{\dag }/q_{n}.$

\textbf{Proof}. At first, one can note that the function $f(\rho
,\sigma )=\mathrm{Tr}\sqrt{\rho }\sqrt{\sigma }$  is closely related to the
QSI and has many useful properties \cite{luo}:

(I) $f(\rho \otimes \tau ,\sigma \otimes \tau )=\mathrm{Tr}\sqrt{\rho }\sqrt{%
\sigma }=f(\rho ,\sigma )$ for any density matrix $\tau $;

(II) $f(U\rho U^{\dag },U\sigma U^{\dag })=f(\rho ,\sigma )$ for any unitary
operation;

(III) (joint concavity) $f(\rho ,\sigma )\leq f(\$\left[ \rho \right] ,\$%
\left[ \sigma \right] )$ for any quantum channel $\$$.

With the above properties, we can begin our proof as follows. Any quantum
channel $\$$ can always be implemented by first utilizing a proper unitary
evolution on the composite system composed of the system of interest and an
auxiliary system and then performing a proper projective measurement on the
auxiliary system, i.e., 
\begin{equation}
M_{n}\rho M_{n}^{\dag }\otimes \left\vert n\right\rangle _{a}\left\langle
n\right\vert =\left\Vert n\right\rangle _{a}\left\langle n\right\Vert
U\left( \rho \otimes \left\vert 0\right\rangle _{a}\left\langle 0\right\vert
\right) U^{\dag }\left\Vert n\right\rangle _{a}\left\langle n\right\Vert ,
\label{dengjia}
\end{equation}%
where $\left\Vert n\right\rangle _{a}=$ $\mathbb{I}\otimes \left\vert
n\right\rangle _{a}$ denotes the orthonormal basis in the auxiliary space
(labelled by $a$), $U$ is a unitary operation on the composite system
determined by $\$$. Explicitly, we have $M_{n}=\left\langle n\right\Vert
_{a}U\left\Vert 0\right\rangle _{a}$.

According to the properties (I) and (II), we have 
\begin{equation}
f(\rho ,\sigma )=f(U\left( \rho \otimes \tau _{a}\right) U^{\dag },U\left(
\sigma \otimes \tau _{a}\right) U^{\dag }).  \label{zj}
\end{equation}%
Let $\tau _{a}=\left\vert 0\right\rangle _{a}\left\langle 0\right\vert $ and 
$\$^{\prime }=\{\left\Vert n\right\rangle _{a}\left\langle n\right\Vert \}$,
then the property (III) and Eq. (\ref{dengjia}) imply 
\begin{eqnarray}
&&f(\rho ,\sigma )\leq f(\$^{\prime }\left[ U\left( \rho \otimes \tau
_{a}\right) U^{\dag }\right] ,\$^{\prime }\left[ U\left( \sigma \otimes \tau
_{a}\right) U^{\dag }\right] )  \notag \\
&=&f(\sum_{n}M_{n}\rho M_{n}^{\dag }\mathbf{\otimes }\left\vert
n\right\rangle _{a}\left\langle n\right\vert ,\sum_{n^{\prime }}M_{n^{\prime
}}\sigma M_{n^{\prime }}^{\dag }\mathbf{\otimes }\left\vert n^{\prime
}\right\rangle _{a}\left\langle n^{\prime }\right\vert )  \notag \\
&=&\sum_{n}f\left( M_{n}\rho M_{n}^{\dag },M_{n}\sigma M_{n}^{\dag }\right)
=\sum_{n}\sqrt{p_{n}q_{n}}f\left( \rho _{n},\sigma _{n}\right) ,  \label{1s}
\end{eqnarray}%
with $p_{n}=$Tr$M_{n}\rho M_{n}^{\dag }$, $q_{n}=$Tr$M_{n}\sigma M_{n}^{\dag
}$ and $\rho _{n}=M_{n}\rho M_{n}^{\dag }/p_{n},\sigma _{n}=M_{n}\sigma
M_{n}^{\dag }/q_{n}$. Here we use the orthonormalization of $\{\left\vert n_a\right\rangle\}$ to 
derive  Eq. (\ref{1s}) which closes the proof.\hfill$\blacksquare$

With Lemma 1 and Lemma 2, now we can prove the theorem 1 as follows.

\textbf{Proof of Theorem 1}. To prove the theorem 1, we need to show the coherence
measure $C\left(\rho\right)$ satisfies all the required criteria (a), (b1),
(b2) and (c).

It is clear that quantum skew information $I\left( \rho ,A\right) $ has many
good properties such as vanishing iff $[\rho ,A]=0$, convexity on the
classical mixing of the states and so on \cite{skew1,skew2,skewl}. $C\left(
\rho \right) $ inherits all the properties, so $C\left( \rho \right) =0$ is
the sufficient and necessary condition for incoherent states and $C\left(
\rho \right) $ is convex under the mixing of states. That is, the criteria
(a) and (c) are automatically satisfied. In addition, one can note that
since the coherence measure is convex, the monotonicity on selective ICPTP
(strong monotonicity) will automatically imply the monotonicity on ICPTP. So
the remaining task of the proof is to prove that $C\left( \rho \right) $
satisfies (b1) -----\textit{the strong monotonicity.}

To do so, let's consider a density matrix $\rho$ with its coherence $%
C\left(\rho\right)$ defined by Eq. (\ref{bianxing}). Meanwhile, we let $\hat{%
\delta}^o$ denote the optimal incoherent state achieving the maximal value
in Eq. (\ref{bianxing}). Define \textit{the incoherent selective quantum
operations}  $\$_{I}$ given by the Kraus operators as $M_{n}$. Suppose $%
\$_{I}$ is performed on the state $\rho $, then the post-measurement
ensemble can be given by $\left\{ p_{n},\rho _{n}\right\} $ with $p_{n}=%
\mathrm{Tr}M_{n}\rho M_{n}^{\dag }$ and $\rho _{n}=M_{n}\rho M_{n}^{\dag
}/p_{n}$. Therefore, the average coherence can be given by%
\begin{equation}
\sum\limits_{n}p_{n}C\left( \rho _{n}\right) =1-\sum\limits_{n}p_{n}\left[
\max_{\hat{\delta}_{n}\in \mathcal{I}}f\left( \frac{M_{n}\rho M_{n}^{\dag }}{%
p_{n}},\hat{\delta}_{n}\right) \right] ^{2}.  \label{tuidaoyi}
\end{equation}%
Since the incoherent operation cannot prepare the coherence from an
incoherent state, for the optimal incoherent state $\hat{\delta}^o$, we have 
$\hat{\delta}_{n}^{o}=\frac{M_{n}\hat{\delta}^{o}M_{n}^{\dag }}{q_{n}}\in $ $%
\mathcal{I}$ with $q_{n}=\mathrm{Tr}M_{n}\hat{\delta}^{o}M_{n}^{\dag } $ for
any incoherent operation $M_{n}$. Thus for such a particular $\hat{\delta}%
_{n}^o$, it is natural that 
\begin{equation}
f\left( \rho _{n},\hat{\delta}_{n}^{o}\right)\leq\max_{\hat{\delta}_{n}\in 
\mathcal{I}}f\left( \rho _{n},\hat{\delta}_{n}\right) .
\end{equation}
Thus Eq. (\ref{tuidaoyi}) can be rewritten as 
\begin{equation}
\sum\limits_{n}p_{n}C\left( \rho _{n}\right) \leq
1-\sum\limits_{n}p_{n}f^{2}\left( \rho _{n},\hat{\delta}_{n}^{o}\right).
\end{equation}
For the probability distribution $\left\{q_n\right\}$, the Cauchy-Schwarz
inequality implies 
\begin{equation}
\sum\limits_{n}p_{n}f^{2}\left( \rho _{n},\hat{\delta}_{n}^{o}\right)\geq %
\left[ \sum\limits_{n}\sqrt{p_{n}q_{n}}f\left( \rho ,\hat{\delta}^{o}\right) %
\right] ^{2}.
\end{equation}
Based on Eq. (\ref{lem2}) given by Lemma 2, we have 
\begin{equation}
\sum\limits_{n}p_{n}C\left( \rho _{n}\right) \leq 1-f^{2}\left( \rho ,\hat{%
\delta}^{o}\right) =C\left( \rho \right)
\end{equation}
which is the strong monotonicity. The convexity of $C\left( \rho \right)$
directly shows $C\left( \rho \right) \geq C\left( \sum_{n=1}p_{n}\rho
_{n}\right) =C\left( \$_{I}\left[ \rho \right] \right)$, that is, the
monotonicity. \hfill $\blacksquare $

\section{Connection with K-coherence for qubits}

The K-coherence of a density matrix $\rho$ subject to a given observable $K$
is defined by \cite{Giro} 
\begin{equation}
C_{K}(\rho )=-\frac{1}{2}\mathrm{Tr}\left\{[\sqrt{\rho },K]\right\}^{2}.
\label{gira}
\end{equation}
Needless to say whether the K-coherence is strongly monotonic or not, it is
obvious that $C_{K}(\rho )$ depends on both the eigenvalue and the
eigenvectors (basis) of $K$. So once the observable $K$ has a degenerate
subspace, the coherence of the state $\rho$ in the corresponding the
subspace won't be revealed. However, our coherence measure $C\left( \rho
\right) $ depends on the broken instead of the original observable, so it is
independent of the eigenvalues of the observable. In other words, it is not
affected by the degeneracy of the observable and so is unambiguously defined
for a certain basis. This is the obvious difference between the K-coherence and ours.
However, next we will show that the K-coherence is only valid for the qubit
system because it is equivalent to our measure $C\left(\rho\right)$
for qubits.

For a \textit{qubit} state $\rho $ and an observable $K$ with the
eigen-decomposition $K=\sum_{k=0}^{1}a_{k}\left\vert k\right\rangle
\left\langle k\right\vert $ where $a_{k}$ is the eigenvalue and $\{\left\vert
k\right\rangle \}$ denotes the set of eigenvectors, our coherence
measure $C\left(\rho \right)$ subject to the basis $\{\left\vert
k\right\rangle \}$ is given by 
\begin{equation}
C(\rho )=-\frac{1}{2}\sum_{k=0}^{1}\mathrm{Tr}\left\{[\sqrt{\rho }%
,\left\vert k\right\rangle \left\langle k\right\vert ]\right\}^{2}
\label{ourd}
\end{equation}
and the K-coherence is given as the same form as Eq. (\ref{gira}). Any
2-dimensional observable can be decomposed as $K=\frac{1}{2}$Tr${K}\cdot 
\mathbb{I}+\tilde{K}$ with $\tilde{K}=\lambda \left( \left\vert
0\right\rangle \left\langle 0\right\vert -\left\vert 1\right\rangle
\left\langle 1\right\vert \right) $ where $\left\vert 0\right\rangle $ and $%
\left\vert 1\right\rangle $ respectively denote the common eigenvectors of $%
K $ and $\tilde{K}$, $\lambda $ represents the positive eigenvalue of $%
\tilde{K}$ and $a_{0/1}$ can be rewritten by $\frac{TrK}{2}\pm \lambda $.
Therefore, Eq. (\ref{gira}) can also be rewritten based on $\tilde{K}$ as 
\begin{eqnarray}
C_{K}(\rho ) &=&-\frac{1}{2}\mathrm{Tr}\left\{\frac{1}{2}\mathrm{Tr}K\left[%
\sqrt{\rho},\mathbb{I}\right]+\left[ \sqrt{\rho },\tilde{K}\right]\right\}
^{2}  \notag \\
&=&-\frac{\lambda ^{2}}{2}\mathrm{Tr}\left\{\left[ \sqrt{\rho },\left\vert
0\right\rangle \left\langle 0\right\vert -\left\vert 1\right\rangle
\left\langle 1\right\vert \right] \right\}^{2}  \notag \\
&=&-\frac{\lambda ^{2}}{2}\left(\frac{1}{2}\mathrm{Tr}\left\{\left[ \sqrt{%
\rho },\mathbb{I}-2\left\vert 1\right\rangle\left\langle 1\right\vert\right]
\right\}^{2}\right.  \notag \\
&&+ \left.\frac{1}{2}\mathrm{Tr}\left\{\left[ \sqrt{%
\rho },2\left\vert 0\right\rangle\left\langle 0\right\vert-\mathbb{I}\right]
\right\}^{2}\right)  \notag \\
&=&2\lambda ^{2}C(\rho ),  \label{eqn17}
\end{eqnarray}%
which exhibits the equivalence between the two coherence measures for qubit
systems if neglecting a constant $2\lambda ^{2}$. Thus $K$-coherence is valid for qubit systems (satisfying the strong monotonicity),  since our
coherence measure $C(\rho)$ is strongly monotonic.

\section{Connection with quantum metrology}

In the following, we will demonstrate how our coherence measure can be
related to some quantum metrology scheme. This also provides an operational
meaning for our coherence measure $C(\rho)$.

The scheme is described as follows. Suppose we have an $n$-dimensional state 
$\rho $ and then let the state undergo a unitary operation $U_{\varphi
_{k}}=e^{-i\varphi _{k}\left\vert k\right\rangle \left\langle k\right\vert }$
which will endow an unknown phase $\varphi _{k}$ to the state $\rho $ as $%
\rho _{k}=U_{\varphi _{k}}\rho U_{\varphi _{k}}^{\dagger }$. We aim to
estimate $\varphi _{k}$ in $\rho _{k}$ by $N>>1$ runs of detection on $\rho
_{k}$. The question is what the measurement precision is.

In the above scheme, the measurement precision of $\varphi _{k}$ is
characterized by the uncertainty of the estimated phase $\varphi _{k}^{est}$
defined by 
\begin{equation}
\delta \varphi _{k}=\left\langle \left( \frac{\varphi _{k}^{est}}{\left\vert
\partial \left\langle \varphi _{k}^{est}\right\rangle /\partial \varphi
_{k}\right\vert }-\varphi _{k}\right) ^{2}\right\rangle ^{1/2}
\end{equation}%
which, for an unbiased estimator, is just the standard deviation \cite%
{fis2,fis1,fis11}. Based on the quantum parameter estimation\cite%
{fis2,fis1,fis11}, $\delta \varphi _{k}$ is limited by the quantum Cram\'{e}%
r-Rao bound as 
\begin{equation}
\left( \delta \varphi _{k}\right) ^{2}\geq \frac{1}{NF_{Qk}},  \label{cra}
\end{equation}%
where $F_{Qk}=\mathrm{Tr}\{\rho _{\varphi }L_{\varphi }^{2}\}$ is the
quantum Fisher information with $L_{\varphi }$ being the symmetric
logarithmic derivative defined by $2\partial _{\varphi }\rho _{\varphi
}=L_{\varphi }\rho _{\varphi }+\rho _{\varphi }L_{\varphi }$ \cite{fis2}. It
was shown in Refs. \cite{fis2,fis1,fis11} that this bound can always be
reached asymptotically by maximum likelihood estimation and a projective
measurement in the eigen-basis of the "symmetric logarithmic derivative
operator" . Thus one can let $\left( \delta \varphi _{k}^{o}\right) ^{2}$ to
denote the optimal variance which achieves the Cram\'{e}r-Rao bound, i.e., $%
\left( \delta \varphi _{k}^{o}\right) ^{2}=\frac{1}{NF_{Qk}}$. Ref. \cite%
{luo1} showed that the Fisher information $F_{Qk}$ is well bounded by the
skew information as 
\begin{equation}
I\left( \rho ,\left\vert k\right\rangle \left\langle k\right\vert \right)
\leq \frac{F_{Qk}}{4}\leq 2I\left( \rho ,\left\vert k\right\rangle
\left\langle k\right\vert \right) ,
\end{equation}%
which directly leads to 
\begin{equation}
4NI(\rho ,\left\vert k\right\rangle \left\langle k\right\vert )\leq \frac{1}{%
\left( \delta \varphi _{k}^{o}\right) ^{2}}\leq 8NI(\rho ,\left\vert
k\right\rangle \left\langle k\right\vert ).  \label{resufi}
\end{equation}%
Suppose we repeat this scheme $N$ times respectively corresponding to
the different $\left\vert k\right\rangle \left\langle k\right\vert $, we can
sum Eq. (\ref{resufi}) over $k$ as 
\begin{equation}
4NC\left( \rho \right) \leq \sum_{k}\frac{1}{\left( \delta \varphi
_{k}^{o}\right) ^{2}}\leq 8NC\left( \rho \right) ,\label{fre}
\end{equation}%
where we have used $C(\rho )=\sum_{k}I(\rho ,\left\vert k\right\rangle
\left\langle k\right\vert )$. If we define $\frac{1}{(\Delta _{\varphi
}^{o})^{2}}=\sum_{k}\frac{1}{\left( \delta \varphi _{k}^{o}\right) ^{2}}$,
Eq. (\ref{fre}) can be rewritten as 
\begin{equation}
\frac{1}{8NC(\rho )}\leq (\Delta _{\varphi }^{o})^{2}\leq \frac{1}{4NC(\rho )%
},
\end{equation}%
which shows that quantum coherence $C(\rho )$ contributes to the upper and
lower bounds of the "average variance" $(\Delta _{\varphi }^{o})^{2}$ that
characterizes the contributions of all the inverse optimal variances of the
estimated phases.

In fact, one can recognize that the practical variance $\delta\varphi_k$
usually deviates from the optimal one $\delta\varphi^o_k$ because the
experimental measurement strategy cannot be as ideal as we expect
theoretically, so that $\delta\varphi_k\geq\delta\varphi^o_k$. Thus, one can
replace $\delta\varphi^o_k$ in Eq. (\ref{resufi}) and Eq. (\ref{fre}) by $%
\delta\varphi_k$ and obtain the other two relations as 
\begin{equation}
\frac{1}{\left( \delta \varphi _{k}\right) ^{2}}\leq 8NI(\rho,\left\vert
k\right\rangle\left\langle k\right\vert)  \label{general1}
\end{equation}%
and 
\begin{equation}
\sum_{k}\frac{1}{\left( \delta \varphi _{k}\right) ^{2}}\leq 8NC\left(
\rho\right).  \label{general2}
\end{equation}
Eqs. (\ref{general1}) and (\ref{general2}) mean that no matter what kind of
measurement strategy is employed, with the fixed $N$ the measurement cannot
be unlimited precise. The variance $\varphi_k$ is well restricted by the
skew information $I(\rho,\left\vert k\right\rangle\left\langle k\right\vert)$
(of course by the corresponding Fisher information), while the sum of $\frac{1}{\varphi_k^2}$ (or the corresponding $\frac{1}{(\Delta _{\varphi
})^{2}}$ ) is just constrained by our coherence $C(\rho)$.

\section{Distribution of coherence}

In this section, we will consider how the coherence is distributed among a
multipartite system. This essentially requires to extend the coherence to
multipartite system and establish the trade-off relation between the
coherence among different subsystems and even the relation with other
quantum features. Such a question was considered by Ref. \cite{Radha}, but
the tradeoff relation as mentioned at the beginning includes both the
basis-free coherence measure and the basis-dependent coherence measure,
especially, this relation depends on the state (monogamous for some states and
polygamous for other states.). This indeed benefits our recognition of
coherence, but strictly speaking, should be the property of the state instead of
the coherence. So how to establish a tradeoff relation
describing a certain property (monogamy or polygamy) with the unified
measure is very important no matter it serves as a physical feature or a
physical resource.  In order to keep the consistent reference basis (similar
to the monogamy of entanglement via the same entanglement quantifier \cite%
{mono1,mono2}), we will restrict ourselves into the computational basis with
which our coherence can be directly used. Therefore, the polygamy relation
of bipartite pure states can be given as follows.

\textbf{Theorem 2}\textit{.}-For a bipartite pure state $\left\vert \Psi
\right\rangle _{AB}$, let $\rho _{A/B}$ denote the reduced density matrix
for $A$ or $B$, then 
\begin{equation}
1-C(\left\vert \Psi \right\rangle _{AB})\leq \lbrack 1-C(\rho
_{A})][1-C(\rho _{B})]  \label{theo2}
\end{equation}%
which is saturated by product states.

\textbf{Proof}. The pure state $\left\vert \Psi \right\rangle _{AB}$ has the
Schmidt decomposition as $\left\vert \Psi \right\rangle
_{AB}=\sum\limits_{i}\lambda _{i}\left\vert \mu _{i}\right\rangle \left\vert
\nu _{i}\right\rangle $ from which we can rewrite $\left\vert \Psi
\right\rangle _{AB}=\sum\limits_{i}\lambda _{i}U_{A}\otimes U_{B}\left\vert
\mu _{i}\right\rangle \left\vert \nu _{i}\right\rangle $ with $\lambda _{i}$
the Schmidt coefficients, so the reduced density matrices can be
respectively given by $\rho _{A}=$ $\sum\limits_{i}\lambda
_{i}^{2}U_{A}\left\vert \mu _{i}\right\rangle \left\langle \mu
_{i}\right\vert U_{A}^{\dag }$ and $\rho _{B}=$ $\sum\limits_{i}\lambda
_{i}^{2}U_{B}\left\vert \nu _{i}\right\rangle \left\langle \nu
_{i}\right\vert U_{B}^{\dag }$. Thus one can always calculate the coherence
for $\left\vert \Psi \right\rangle _{AB}$ and its reduced matrices $\rho
_{A} $ and $\rho _{B}$ (within the basis $\left\vert k\right\rangle $ and $%
\left\vert k^{\prime }\right\rangle $ instead of the Schmidt basis $%
\left\vert \mu _{i}\right\rangle $ and $\left\vert \nu _{i}\right\rangle $)
as 
\begin{eqnarray}
1-C\left( \left\vert \Psi \right\rangle _{AB}\right)
&=&\sum\limits_{kk^{\prime }}\left\vert \sum\limits_{i}\lambda
_{i}\left\langle k\right\vert U_{A}\left\vert \mu _{i}\right\rangle
\left\langle k^{\prime }\right\vert U_{B}\left\vert \nu _{i}\right\rangle
\right\vert ^{4}, \\
1-C\left( \rho _{A}\right) &=&\sum\limits_{k}\left[ \sum\limits_{i}\lambda
_{i}\left\vert \left\langle k\right\vert U_{A}\left\vert \mu
_{i}\right\rangle \right\vert ^{2}\right] ^{2}, \\
1-C\left( \rho _{B}\right)& =&\sum\limits_{k^{\prime }}\left[
\sum\limits_{i}\lambda _{i}\left\vert \left\langle k^{\prime }\right\vert
U_{B}\left\vert \nu _{i}\right\rangle \right\vert ^{2}\right] ^{2}.
\end{eqnarray}%
From these three equations, we can find that for each $k$ and $k^{\prime }$, 
\begin{eqnarray}
&&\left( \sum\limits_{i}\lambda _{i}\left\vert \left\langle k\right\vert
U_{A}\left\vert \mu _{i}\right\rangle \right\vert ^{2}\right) \cdot \left(
\sum\limits_{i}\lambda _{i}\left\vert \left\langle k^{\prime }\right\vert
U_{B}\left\vert \nu _{i}\right\rangle \right\vert ^{2}\right)  \notag \\
&\geqslant &\left( \sum\limits_{i}\lambda _{i}\left\vert \left\langle
k\right\vert U_{A}\left\vert \mu _{i}\right\rangle \right\vert \cdot
\left\vert \left\langle k^{\prime }\right\vert U_{B}\left\vert \nu
_{i}\right\rangle \right\vert \right) ^{2}  \notag \\
&\geq &\left\vert \sum\limits_{i}\lambda _{i}\left\langle k\right\vert
U_{A}\left\vert \mu _{i}\right\rangle \left\langle k^{\prime }\right\vert
U_{B}\left\vert \nu _{i}\right\rangle \right\vert ^{2}.  \label{iminn}
\end{eqnarray}%
Therefore, squaring both sides of Eq. (\ref{iminn}) and summing over $k$ and 
$k^{\prime }$, one will immediately arrive at Eq. (\ref{theo2}). It is easy
to show that the product states saturate the inequality.\hfill $\blacksquare 
$

From theorem 2, it can be found that the coherence of a subsystem is not
limited by the coherence of the composite system. A trivial case is that the
incoherent composite quantum state means no coherence in its subsystems.
However, the composite quantum state with the relatively large coherence
doesn't restrict the coherence of the subsystems (which is different from
the monogamy of entanglement). That is, the subsystems could also have the
relatively large coherence. A typical example is the maximally coherent
state, e.g. $\left\vert \Psi \right\rangle _{AB}=\frac{1}{3}%
\sum_{i,j=0}^{2}\left\vert ij\right\rangle $. One can find that $C\left(
\left\vert \Psi \right\rangle _{AB}\right) =\frac{8}{9}$ but $C\left( \rho
_{A}\right) =C\left( \rho _{B}\right) =\frac{2}{3}$ which is the maximal
coherence in 3-dimensional space corresponding to the reduced states $\rho
_{A}=\rho _{B}=\frac{1}{3}\sum_{i,j=0}^{2}\left\vert i\right\rangle
\left\langle j\right\vert $. This example also implies that the subsystem
with the relatively large coherence doesn't restrict its ability to interact
with another system and form a composite system with the large coherence.
These are the manifestation of the so-called \textit{polygamy}. Theorem 2
can also be extended to mixed states and multipartite states as the
following two corollaries.

\textbf{Corollary 1}.- For bipartite mixed states $\rho _{AB}$ with its
reduced density matrices $\rho _{A/B}$, the coherences satisfy 
\begin{gather}
\lbrack 1-C(\rho _{A})][1-C(\rho _{B})]\geq \sum_{kk^{\prime }}\left\langle
kk^{\prime }\right\vert \rho _{AB}\left\vert kk^{\prime }\right\rangle ^{2}
\label{mon1} \\
=\mathrm{Tr}\rho _{AB}^{2}-C_{2}(\rho _{AB})\geq \lambda _{\min }\left[
1-C(\rho _{AB})\right]  \label{mon1p}
\end{gather}%
with $\left\vert kk^{\prime }\right\rangle $ being the fixed computational
basis, $\lambda _{\min }$ denoting the minimal \textit{nonzero} eigenvalue
of $\rho _{AB}$ and $C_{l_k}\left( \rho \right)$ denoting the $l_k$-norm
coherence. In addition, one can also have 
\begin{equation}
\lbrack 1-C(\rho _{A})]\left[ r-\sum\limits_{i=1}^{r}C\left( \rho
_{Bi}\right) \right] \geqslant 1-C(\rho _{AB}),  \label{mon2}
\end{equation}%
\begin{equation}
\lbrack r-\sum\limits_{i=1}^{r}C(\rho _{Ai})]\left[ 1-C\left( \rho
_{B}\right) \right] \geqslant 1-C(\rho _{AB}),  \label{mon3}
\end{equation}%
which can also lead to a symmetric form as%
\begin{equation}
\lbrack 1-C(\rho _{A})][1-C(\rho _{B})]\geq \frac{1}{c_{s}}\left[ 1-C(\rho
_{AB})\right] ^{2}  \label{monsym}
\end{equation}%
with $c_{s}=[r-\sum\limits_{i}C(\rho _{Ai})][r-\sum\limits_{i}C(\rho _{Bi})]$
where $r$ is the rank of $\rho _{AB}$ and $\rho _{Ai}$, $\rho _{Bi}$ denote
the reduced density matrices of $i$th eigenstate of $\rho _{AB}$.

\textbf{Corollary 2}.- For an $N$-partite quantum state $\rho _{AB\cdots N}$%
, define the index set $S=\left\{ A,B,C,\cdots ,N\right\} $ corresponding to
all the $N$ subsystems. Let $\alpha $ represent a subset of $S$, i.e., $%
\alpha \subset S$ and $\rho _{\alpha }$ denote the reduced density matrix by
tracing over all subsystems corresponding to $\bar{\alpha}$, the
complementary set of $S$. Thus for $\forall \alpha _{i}\subset S$ such that $%
\alpha _{i}\cap \alpha _{j}=\delta _{ij}\alpha _{i}$ and $\sum_{i=1}\alpha
_{i}=S$, the coherences satisfy 
\begin{eqnarray}
\prod\limits_{i}\left[ 1-C(\rho _{\alpha _{i}})\right] &\geq &\lambda _{M}%
\left[ 1-C(\rho _{AB\cdots N})\right] ,  \label{mont} \\
\prod\limits_{i}\left[ 1-C(\rho _{\alpha _{i}})\right] ^{n_{i}} &\geq &\frac{%
1}{c_{sT}}\left[ 1-C(\rho _{AB\cdots N})\right] ^{2},  \label{monst}
\end{eqnarray}%
where $n_{i}$ as well as $\lambda _{M}$ and $c_{sT}$ can be determined from
Corollary 1 based on the concrete bipartite grouping of $\rho _{AB\cdots N}.$

The proofs of both Corollary 1 and Corollary 2 are given in the Appendix \ref{app:poly}
which also demonstrates how to determine $n_{i}$, $\lambda _{M}$ and $c_{sT}$%
. One can note that Eq. (\ref{mon1p}) can be understood as the general
polygamy relation for both mixed and pure states since $\lambda_{\min}=1$
for pure state. In addition, no matter what $\lambda _{M},c_{sT},\lambda
_{\min }$ and $c_{s}$ are, they can always be some finite values. Therefore,
similar to theorem 2, the \textit{polygamy} is also clearly demonstrated by
mixed states and multipartite states.

\section{Bounds on quantum discord}

The resource theory provides a platform to understand one quantum feature
via another quantum feature. Quantum coherence can be understood by quantum
discord \cite{Ma}. That is, the coherence assisted by an incoherent
auxiliary state can be converted by incoherent operations to the same amount
of quantum discord. As an application of our coherence measure, here we
revisit this question and find some similar bounds. As we know, quantum
discord of a bipartite quantum state is initially defined by the discrepancy
between quantum versions of two classically equivalent expressions for
mutual information \cite{disc1,disc2}. Even though the latter various
measures of quantum discord have been presented \cite{review}, quantum
discord with both the good computability and the good properties (e.g.
contractivity) should count on local quantum uncertainty (LQU) based on
quantum skew information \cite{lqu}. We would like to emphasize that the LQU
was developed with the broken observable in Ref. \cite{Yuquan}. In the following, we will restrict the
quantum discord  to the one given in Ref. \cite{Yuquan}.

The quantum discord in Ref. \cite{Yuquan} is defined for a bipartite state $%
\rho _{AB}$ as 
\begin{equation}
D\left( \rho _{AB}\right) =\min_{\left\{ \left\vert k\right\rangle
_{A}\right\} }C_{\left\{ \left\vert k\right\rangle _{A}\right\} }\left( \rho
_{AB}\right),
\end{equation}
where 
\begin{equation}
C_{\left\{ \left\vert k\right\rangle _{A}\right\} }\left( \rho _{AB}\right)
=-\frac{1}{2}\sum_{k}\mathrm{Tr}[\sqrt{\rho _{AB}},\left\vert k\right\rangle
_{A}\left\langle k\right\vert \otimes \mathbb{I}_{B}]^{2}
\end{equation}
and $\left\{ \left\vert k\right\rangle _{A}\right\} $ denotes the fixed
basis. We can understand $C_{\left\{ \left\vert k\right\rangle _{A}\right\}
}\left( \rho _{AB}\right) $ as the coherence of the \textit{A} subspace and
thus $D\left( \rho _{AB}\right) $ can be naturally considered as the minimal
coherence of \textit{A} subspace. Since $I(\rho _{AB},K\otimes \mathbb{I}%
_{B})\geq I(\rho _{A},K)$, one can immediately obtain 
\begin{equation}
C_{\left\{ \left\vert k\right\rangle _{A}\right\} }\left( \rho _{AB}\right)
\geq D\left( \rho _{AB}\right) \geq C_{\left\{ \left\vert \tilde{k}%
\right\rangle _{A}\right\} }\left( \rho _{A}\right)
\end{equation}%
with $\left\{ \left\vert \tilde{k}\right\rangle _{A}\right\} $ denoting the
optimal basis to achieve the quantum discord. This relation implies the
quantum discord is upper bounded by its subspace coherence and lower bounded
by the coherence of the subsystem subject to the optimal basis. To reveal
all the quantum discords, the symmetric quantum discord can be similarly
defined as 
\begin{equation}
D_{S}\left( \rho _{AB}\right) =\min_{\left\{ \left\vert k\right\rangle
\right\} \left\{ \left\vert k^{\prime }\right\rangle \right\} }C_{\left\{
\left\vert kk^{\prime }\right\rangle \right\} }\left( \rho _{AB}\right)
\label{sdiscord}
\end{equation}%
with 
\begin{equation}
C_{\left\{ \left\vert kk^{\prime }\right\rangle \right\} }\left( \rho
_{AB}\right) =-\frac{1}{2}\sum_{kk^{\prime }}\mathrm{Tr}[\sqrt{\rho _{AB}}%
,\left\vert k\right\rangle _{A}\left\langle k\right\vert \otimes \left\vert
k^{\prime }\right\rangle _{B}\left\langle k^{\prime }\right\vert ]^{2}.
\end{equation}
Analogously, $C_{\{\left\vert kk^{\prime }\right\rangle\} }\left( \rho
_{AB}\right) $ is exactly the coherence of $\rho _{AB}$ within the basis $%
\left\{ \left\vert k\right\rangle \left\vert k^{\prime }\right\rangle
\right\} $ and quantum discord $D_{S}\left( \rho _{AB}\right) $ is just the
minimal coherence. With these concepts in mind, we can give the important
results in the following rigorous way.

\textbf{Theorem 3}.- Suppose an incoherent operation $\$_I$ is performed on
a bipartite product state $\sigma _{A}\otimes \sigma _{B}$ is a bipartite
product state. The quantum discord of the post-operation state is bounded as 
\begin{equation}
D_{S}\left( \$_{I}[\sigma _{A}\otimes \sigma _{B}]\right) \leq 1-\left(
1-C\left( \sigma _{A}\right) \right) \left( 1-C\left( \sigma _{B}\right)
\right) .  \label{disccc}
\end{equation}
In particular, the upper bound is attained by $\$_I=\left\{U_{I}=\sum_{ij}%
\left\vert i,i\oplus j\right\rangle \left\langle i,j\right\vert\right\}$ and 
$\sigma _{B/A}=\left\vert
k\right\rangle \left\langle k\right\vert $.

\textbf{Proof}. From Eq. (\ref{sdiscord}), one can find that the discord is
gotten by the minimization among all the potential basis, so it is natural
that 
\begin{equation}
D_{S}\left( \$_{I}[\sigma _{A}\otimes \sigma _{B}]\right) \leq C\left(
\$_{I}[\sigma _{A}\otimes \sigma _{B}]\right).
\end{equation}
Based on the monotonicity of the coherence, one will immediately arrive
at 
\begin{eqnarray}
&&C\left(\$_{I}[\sigma _{A}\otimes \sigma _{B}]\right)\leq C\left( \sigma
_{A}\otimes \sigma _{B}\right)  \notag \\
&&=1-\left( 1-C\left( \sigma _{A}\right) \right) \left( 1-C\left( \sigma
_{B}\right) \right) ,  \label{discoha}
\end{eqnarray}%
which shows Eq. (\ref{disccc}) is valid.

Next, we will show the upper bound is attainable as mentioned in the
theorem. Let $\sigma _{B}=\left\vert \tilde{k}\right\rangle \left\langle 
\tilde{k}\right\vert $, so the initial state can be written as $\rho
_{0}=\rho _{A}\otimes \left\vert \tilde{k}\right\rangle \left\langle \tilde{k%
}\right\vert $. Suppose we employ the incoherence operation $\$_{I}=\left\{
U_{I}=\sum_{ij}\left\vert i,i\oplus j\right\rangle \left\langle
i,j\right\vert \right\} $. So the state after the operation is written by $%
\rho _{f}=U_{I}\rho _{0}U_{I}^{\dagger }$. Consider the eigen-decomposition
of $\rho _{A}=\sum_{i}\lambda _{i}\left\vert \psi _{i}\right\rangle
_{A}\left\langle \psi _{i}\right\vert $ with the eigenstate $\left\vert \psi
_{i}\right\rangle =\sum_{j}a_{j}^{i}\left\vert j\right\rangle $ expanded by
the basis $\left\{\left\vert j\right\rangle\right\} $, we can rewrite $\rho _{f}$ as 
\begin{eqnarray}
\rho _{f} &=&\sum_{i}\lambda _{i}U_{I}\left\vert \psi _{i}\right\rangle
_{A}\left\vert \tilde{k}\right\rangle _{B}\left\langle \psi _{i}\right\vert
_{A}\left\langle \tilde{k}\right\vert _{B}U_{I}^{\dagger }  \notag \\
&=&\sum_{i}\lambda _{i}\left( \sum_{j}a_{j}^{i}\left\vert jj\oplus \tilde{k}%
\right\rangle \right) \left( \sum_{j}\left\langle jj\oplus \tilde{k}%
\right\vert a_{j}^{i\ast }\right) .
\end{eqnarray}%
Based on our definition of quantum coherence, we can easily obtain the
quantum coherence of $\rho _{A}$ within the basis $\left\{ \left\vert
j\right\rangle \right\} $ as 
\begin{eqnarray}
C\left( \rho _{A}\right)  &=&1-\sum_{j}\left( \sum_{i}\sqrt{\lambda _{i}}%
\left\vert \left\langle j\right\vert \left. \psi _{i}\right\rangle
\right\vert ^{2}\right) ^{2}  \notag \\
&=&1-\sum_{j}\left( \sum_{i}\sqrt{\lambda _{i}}\left\vert
a_{j}^{i}\right\vert ^{2}\right) ^{2}.  \label{inic}
\end{eqnarray}%
According to the definition of quantum correlation $D_{S}(\cdot )$, one can
find that%
\begin{eqnarray}
&&1-D_{S}\left( \rho _{f}\right)   \notag \\
&=&\max_{\{\left\vert kk^{\prime }\right\rangle \}}\sum_{kk^{\prime }}\left[
\sum_{i}\sqrt{\lambda _{i}}\left( \sum_{j}a_{j}^{i}\left\langle kk^{\prime
}\right. \left\vert jj\oplus \tilde{k}\right\rangle \right) \right.   \notag
\\
&\times &\left. \left( \sum_{j}\left\langle jj\oplus \tilde{k}\right\vert
\left. kk^{\prime }\right\rangle a_{j}^{i\ast }\right) \right] ^{2}  \notag
\\
&=&\max_{\{\left\vert kk^{\prime }\right\rangle \}}\sum_{kk^{\prime }}\left(
\sum_{i}\sqrt{\lambda _{i}}\left\langle kk^{\prime }\right\vert P_{\tilde{k}%
}\Lambda _{i}\otimes \mathbf{1}\left\vert \Phi \right\rangle \left\langle
\Phi \right\vert P_{\tilde{k}}\Lambda _{i}^{\ast }\otimes \mathbf{1}%
\left\vert kk^{\prime }\right\rangle \right) ^{2}  \notag \\
&=&\max_{U,V}\sum_{j}\left( \sum_{i}\sqrt{\lambda _{i}}\left\vert
\left\langle j\right\vert U^{\dag }P_{\tilde{k}}\Lambda _{i}P_{\tilde{k}%
}V^{\ast }\left\vert j\right\rangle \right\vert ^{2}\right) ^{2}  \notag \\
&=&\max_{U,V}\sum_{j}\left( \sum_{i}\sqrt{\lambda _{i}}\left\vert \sum_{k}%
\left[ U^{\dag }\right] _{jk}a_{k}^{i}\left[ V^{\ast }\right]
_{kj}\right\vert ^{2}\right) ^{2}.  \label{zhonj}
\end{eqnarray}%
Here we first use the fact $\sum_{j}a_{j}^{i}\left\vert jj\oplus \tilde{k}%
\right\rangle =\left( P_{\tilde{k}}\Lambda _{i}\otimes \mathbb{I}\right)
\left\vert \Phi \right\rangle $, where $\left\vert \Phi \right\rangle
=\sum_{j}\left\vert jj\right\rangle $, $\Lambda _{i}=diag(a_{0},a_{1},\cdots
)$ and $P_{\tilde{k}}=\sum_{j}\left\vert \tilde{k}\oplus j\right\rangle
\left\langle j\right\vert $. In addition, we also convert the optimization
on the basis $\{\left\vert kk^{\prime }\right\rangle \}$ to the unitary
transformations by $\left\vert k\right\rangle =U\left\vert j\right\rangle $
and $\left\vert k^{\prime }\right\rangle =V\left\vert j\right\rangle $. In
the last line of Eq. (\ref{zhonj}), we omit $P_{\tilde{k}}$ because we force 
$P_{\tilde{k}}$ to be absorbed by the optimized unitary transformations $U$
and $V$.  By utilizing the Cauchy-Schwarz inequality to Eq. (\ref{zhonj}),
one will find 
\begin{eqnarray}
D_{S}\left( \rho _{f}\right)  &\geqslant &1-\max_{U}\sum_{j}\left( \sum_{i}%
\sqrt{\lambda _{i}}\sum_{k}\left\vert \left[ U^{\dag }\right]
_{jk}\right\vert ^{2}\left\vert a_{k}^{i}\right\vert ^{2}\right) ^{2}  \notag
\\
&\geqslant &1-\max_{U}\sum_{jk}\left\vert \left[ U^{\dag }\right]
_{jk}\right\vert ^{2}\left( \sum_{i}\sqrt{\lambda _{i}}\left\vert
a_{k}^{i}\right\vert ^{2}\right) ^{2}  \label{ineq2} \\
&=&1-\sum_{j}\left( \sum_{i}\sqrt{\lambda _{i}}\left\vert
a_{j}^{i}\right\vert ^{2}\right) ^{2},  \label{proofend}
\end{eqnarray}
where the inequality (\ref{ineq2}) comes from the the convexity and the
extreme value is achieved when we select the optimal basis $\left\{
\left\vert kk^{\prime }\right\rangle \right\} =\left\{ \left\vert
jj\right\rangle \right\} $. Comparing Eq. (\ref{proofend}) and Eq. (\ref%
{inic}), one can find 
\begin{equation}
D_{S}\left( \rho _{f}\right) \geqslant C\left( \rho _{A}\right) .
\label{invers}
\end{equation}

\bigskip However, based on Eq. (\ref{disccc}), we have $D_{S}\left( \rho
_{f}\right) \leq C\left( \rho _{A}\right) $ for $\sigma _{B}=\left\vert 
\tilde{k}\right\rangle \left\langle \tilde{k}\right\vert $ and $U_{I}$. This
means in this case $D_{S}\left( \rho _{f}\right) =C\left( \rho _{A}\right) $
which completes the proof. \hfill $\blacksquare $

In fact, if both $\sigma _{A}$ and $\sigma _{B}$ are coherent, one can find
that the upper bound could not be attained generally for the fixed dimension
of the state space. For example, $\sigma _{A}=\sigma _{B}=\frac{1}{2}\left(
\left\vert 0\right\rangle +\left\vert 1\right\rangle \right) \left(
\left\langle 0\right\vert +\left\langle 1\right\vert \right) $, a simple
algebra can show $C\left( \sigma _{A}\otimes \sigma _{B}\right) =\frac{3}{4}$%
, but the maximal quantum discord in this fixed space is $D_{S}\left(
\$_{I}[\sigma _{A}\otimes \sigma _{B}]\right) =\frac{1}{2}$ where $\$_{I}=[%
\mathbb{I}_{2}\oplus i\sigma _{y}]$, $\mathbb{I}_{2}$ and $\sigma _{y}$ are
respectively the 2-dimensional identity matrix and Pauli matrix. However, if
the state space is not fixed, the upper bound is obviously attainable,
because one can always expand the state space as $\sigma _{A/B}\oplus 0$ as
required, which, in some cases, is equivalent to attaching an auxiliary
system as $\sigma _{A}\otimes \sigma _{B}\otimes \left\vert 0\right\rangle
_{C}\left\langle 0\right\vert .$ In this sense, it is apparent that the
coherence of $\sigma _{A}\otimes \sigma _{B}$ can be completely converted to
the quantum discord between $(AB)$ and $C$. One can perform a (incoherent)
swapping operation on $A$ and $C$ and finally obtain the equal amount of
quantum discord between $A$ and $\left( BC\right) $ ($BC$ can be replaced by 
$B$ with the equally expanded space). Finally we would like to emphasize
that the similar Eq. (\ref{discoha}) is also satisfied for multipartite
states.

\section{Directly measurable coherence}

In this section, we will discuss the measurement of coherence in practical
experiments. Like entanglement measure, the coherence measure \textit{per se}
is not an observable. In order to avoid so much cost (mainly in high
dimensional system) for QST, the schemes for the direct measurement of
entanglement and quantum discord have been presented in recent years by the
simultaneous copies of the state \cite{en1,en2,en3,dis1,dis2} or by an
auxiliary system \cite{en4}, which provides a valuable reference for the
coherence measure. For example, the relative-entropy coherence for an $N_{D}$%
-dimensional state $\rho $ is given explicitly by 
\begin{equation}
C_{r}\left( \rho \right) =\sum_{i}\lambda _{i}\log {\lambda _{i}}%
-\sum_{k}\rho _{kk}\log {\rho _{kk}}
\end{equation}
with $\lambda _{i}$'s denoting the eigenvalues of $\rho $ and $\rho
_{kk}=\left\langle k\right\vert \rho \left\vert k\right\rangle $ being the
diagonal entries subject to the basis $\{\left\vert k\right\rangle \}$.
Since $\lambda _{i}$'s can be measured by the standard overlap measurement 
\cite{en4,overlap} and $\rho _{kk}$ can be measured by the given projectors $%
\hat{P}_{k}=\left\vert k\right\rangle \left\langle k\right\vert $, $%
C_{r}\left( \rho \right) $ is experimentally measurable. The cost is $%
2(N_{D}-1)$ measurements assisted by at most $N_{D}$ copies of the state.
The detailed measurement scheme is described for clarity in the Appendix \ref{app:measure}.

In fact, the measurable evaluation of coherence (instead of the exact value
as given above for the relative-entropy coherence) with less cost is also
quite practical. We find that our $C\left( \rho \right) $ can also be
effectively evaluated by the measurable upper and lower bounds. Based on the
inequality $I(A,\rho )\geqslant -\frac{1}{4}\mathrm{Tr}\{[\rho ,A]^{2}\}$
for any observable $A$ and a density matrix $\rho $ \cite{Giro}, we have 

\begin{eqnarray}
C\left( \rho \right)  &=&\sum_{k}I(\left\vert k\right\rangle \left\langle
k\right\vert ,\rho )  \notag \\
&\geqslant &\frac{1}{2}\left( \mathrm{Tr}\rho ^{2}-\sum_{k}\left\langle
k\right\vert \rho \left\vert k\right\rangle ^{2}\right) =\frac{1}{2}%
C_{l_{2}}(\rho )  \label{mlb}
\end{eqnarray}%
with $\left\{\left\vert k\right\rangle \right\}$ defining the basis. Here
 \begin{eqnarray}
 &&C_{l_{2}}\left(
\rho \right) =\left\Vert \rho -\delta _{I}\right\Vert _{2}=\sum_{i\neq
j}\left\vert \rho _{ij}\right\vert ^{2}\notag\\
&=&\mathrm{Tr}\rho ^{2}-\sum_{k}\left\langle
k\right\vert \rho \left\vert k\right\rangle ^{2}
=\sum_k\left\{\lambda_k^2-\left\langle
k\right\vert \rho \left\vert k\right\rangle ^{2}\right\}, \label{57}
\end{eqnarray} where $\left\Vert \cdot \right\Vert
_{2}$ denotes the $l_{2}$ norm of a matrix, $\delta
_{I}=\sum\limits_{k}\rho _{kk}\left\vert k\right\rangle \left\langle
k\right\vert $ is the closest incoherent state and $\lambda_k$'s are the eigenvalues of $\rho$.  In addition, one can also
find that $\left\langle k\right\vert \sqrt{\rho }\left\vert k\right\rangle
\geq \left\langle k\right\vert {\rho }\left\vert k\right\rangle $ is
satisfied for any $\left\vert k\right\rangle $. Thus one can have 
\begin{equation}
C\left( \rho \right) =1-\sum_{k}\left\langle k\right\vert \sqrt{\rho }%
\left\vert k\right\rangle ^{2}\leq 1-\sum_{k}\left\langle k\right\vert {\rho 
}\left\vert k\right\rangle ^{2}.  \label{mub}
\end{equation}%
Combine Eq. (\ref{mlb}) and (\ref{mub}), one will immediately obtain our
second result: 
\begin{equation}
\frac{1}{2}C_{l_{2}}\left( \rho \right) \leq C\left( \rho \right) \leq 1-%
\mathrm{Tr}\rho ^{2}+C_{l_{2}}\left( \rho \right)   \label{mbb}
\end{equation}%
which provides both the upper and the lower bounds. Even though the
coherence based on the $l_{2}$ norm is not a good measure, as one bound, it
serves as a sufficient and necessary condition for the existence of quantum
coherence. Since $C_{l_{2}}$ is completely characterized by the eigenvalues $\lambda_k$ and the diagonal entries $\left\langle k\right\vert\rho\left\vert k\right\rangle$ as seen from Eq. (\ref{57}), one can find that both bounds are practically
measurable similar to the above measurement scheme for the relative-entropy
coherence. The cost is $N_{D}$ measurements plus 2 copies of the state $\rho 
$. 

In fact, $l_{1}$-norm coherence has also the similar measurable bounds. As
we know, for the $N_{D}$-dimensional density matrix $\rho $, we have 
\begin{equation}
C_{l_{1}}(\rho )=\sum_{i\neq j}\left\vert \rho _{ij}\right\vert =\frac{1}{2}%
\sum_{i<j}\left\vert \rho _{ij}\right\vert. 
\end{equation}%
Since $\left\vert \rho _{ij}\right\vert \leq 1$, we have $\left\vert \rho
_{ij}\right\vert ^{2}\leq \left\vert \rho _{ij}\right\vert $ which leads to 
\begin{equation}
C_{l_{1}}(\rho )\geq \frac{1%
}{2}\sum_{i<j}\left\vert \rho _{ij}\right\vert ^{2}=C_{l_{2}}(\rho ).  \label{lbl1}
\end{equation}%
Furthermore, the inequality $\left( \sum_{k=1}^{N_{D}}a_{k}\right) ^{2}\leq
N_{D}\sum_{k=1}^{N_{D}}a_{k}^{2}$ for postive $a_{k}$ directly implies that 
\begin{equation}
C_{l_{1}}(\rho )\leq \sqrt{N_{D}(N_{D}-1)C_{l_{2}}(\rho )}.  \label{ubl2}
\end{equation}%
Combining Eqs. (\ref{lbl1}) and (\ref{ubl2}) give the bounds for $%
C_{l_{1}}(\rho )$ as%
\begin{equation}
C_{l_{2}}(\rho )\leq C_{l_{1}}(\rho )\leq \sqrt{N_{D}(N_{D}-1)C_{l_{2}}(\rho
)}.  \label{l1b}
\end{equation}%
Since $C_{l_{2}}\left( \rho \right) $ is measurable, the above bounds are
naturally measurable. 
In addition, Ref. \cite{Napoli} also proposed a similar lower bound through the ROC and the improved lower bound rather than the exact coherence conditioned on the prior knowledge of the state of interest.

\section{Discussion and Conclusions}

Before the end, we would like to first emphasize that the polygamy inequality shown in theorem 2
has an elegant form for bipartite pure states, but the relation with the
same form doesn't hold for a general bipartite mixed state of qubits, even
though Eq. (14) provides a general polygamy relation. However, we would like
to conjecture that it could hold for the bipartite mixed states with the
dimension $N \geq 6$. The details can be seen from the Appendix \ref{app:conj}.

In summary, we have presented a strongly monotonic coherence measure in terms of quantum skew information which 
characterizes the contribution of the commutation between the broken observable (basis) and the density matrix of interest. It is shown
that the coherence measure has an operational meaning based on the quantum metrology.  We also study the distribution of the coherence among a multipartite system by providing the polygamy inequalities and find 
that the coherence can serve as the natural upper bound on the quantum discord. Finally, we find that our coherence measure as well as the $l_1$-norm can induce the experimentally measurable bounds of coherence, but the relative-entropy coherence can be in principle exactly measured in experiment. 

\section{Acknowledgements}

Yu thanks A. Winter and M. Nath Bera for valuable discussions. This work was
supported by the National Natural Science Foundation of China, under Grant
No.11375036, the Xinghai Scholar Cultivation Plan and the Fundamental
Research Funds for the Central Universities under Grant No. DUT15LK35 and
No. DUT15TD47.

\appendix

\section{An example for K-coherence violating the (strong) monotonicity}\label{app:mono}
Ref. \cite{Giro} defined the $K$-coherence of a state subject to the
observable $K$ by the quantum skew information instead of the direct
commutation. That is, 
\begin{equation}
C_{K}\left( \rho \right) =-\frac{1}{2}Tr[\sqrt{\rho },K]^{2}.  \label{prl1}
\end{equation}%
However, the quantification of coherence given in Eq. (1) not only includes
the contribution of the basis which the observable defines, but also
includes the contribution of the eigenvalues of the observable. In
particular, once the observable is degenerate, the observable won't extract
all the coherence of the state, even though it should be valid in its own
right. The most important is that such a definition only serves as a good
coherence measure in qubit system which will be shown in the following
section. One can easily find that in the general case, this coherence
measure satisfies neither the criterion (b1) nor (b2) in the main text. So
it is not a good coherence measure in general cases, which is also found in
Ref. \cite{Marvian2}. To see this, let's consider the state 
\begin{equation}
\rho =\left( 
\begin{array}{ccc}
0.6309 & 0.0359 & 0.0858 \\ 
0.0359 & 0.0441 & 0.1189 \\ 
0.0858 & 0.1189 & 0.3250%
\end{array}%
\right)
\end{equation}%
undergoes the incoherent quantum channel $\$_{I}=\{M_{n}\}$ with $%
M_{1}=\left( 
\begin{array}{ccc}
0 & 0.3 & 0 \\ 
0 & 0 & 0.5 \\ 
0.7 & 0 & 0%
\end{array}%
\right) $ and $M_{2}=\left( 
\begin{array}{ccc}
0 & 0 & 0.8660 \\ 
0 & 0.9539 & 0 \\ 
0.7141 & 0 & 0%
\end{array}%
\right) $ and $M_{1}^{\dagger }M_{1}+M_{2}^{\dagger }M_{2}=\mathbb{I}_{3}$.
One can obtain the state $\rho _{1}=M_{1}\rho M_{1}^{\dagger }/p_{1}$ with
the probability $p_{1}=TrM_{1}\rho M_{1}^{\dag }$ and the state $\rho
_{2}=M_{2}\rho M_{2}^{\dagger }/p_{2}$ with the probability $%
p_{2}=TrM_{2}\rho M_{2}^{\dagger }$. It is easy to find that the average
coherence $\bar{C}_{K}=p_{1}C_{K}(\rho _{1})+p_{2}C_{K}(\rho _{2})=1.2928$
and the coherence $C_{K}(\rho ^{\prime })$ of the final state $\rho ^{\prime
}=p_{1}\rho _{1}+p_{2}\rho _{2}$ is given by $C_{K}(\rho ^{\prime })=0.3350$%
, while the coherence of the initial state $C_{K}(\rho )=0.2277$ where the
reference observable $K=\left( 
\begin{array}{ccc}
1 & 0 & 0 \\ 
0 & 7 & 0 \\ 
0 & 0 & 5%
\end{array}%
\right) $. It is apparent that the criteria (b1) and (b2) are simultaneously
violated.

\section{Proof of the polygamy of our coherence}\label{app:poly}
\subsection{Proof of Corollary 1}
From the proof of theorem 2, one can find that 
\begin{equation}
\left\langle k\right\vert \sqrt{\rho _{A}}\left\vert k\right\rangle
\left\langle k^{\prime }\right\vert \sqrt{\rho _{B}}\left\vert k^{\prime
}\right\rangle \geq \left\langle kk^{\prime }\right\vert \left. \Psi
\right\rangle _{AB}\left\langle \Psi \right. \left\vert kk^{\prime
}\right\rangle  \label{iminr}
\end{equation}%
holds for pure $\left\vert \Psi \right\rangle _{AB}.$Consider a mixed state
with a potential decomposition $\rho _{AB}=\sum_{i}p_{i}\left\vert \psi
_{i}\right\rangle _{AB}\left\langle \psi _{i}\right\vert $ and substitute
every $\left\vert \psi _{i}\right\rangle _{AB}$ into Eq. (\ref{iminr}), one
will arrive at 
\begin{equation}
\sum_{i}p_{i}\left\langle k\right\vert \sqrt{\rho _{Ai}}\left\vert
k\right\rangle \left\langle k^{\prime }\right\vert \sqrt{\rho _{Bi}}%
\left\vert k^{\prime }\right\rangle \geq \sum\limits_{i}p_{i}\left\langle
kk^{\prime }\right\vert \left. \psi _{i}\right\rangle _{AB}\left\langle \psi
_{i}\right. \left\vert kk^{\prime }\right\rangle .  \label{medir}
\end{equation}%
Squaring both sides of Eq. (\ref{medir}) and summing over all the $%
kk^{\prime }$, we have
\begin{eqnarray}
&&\sum_{kk^{\prime }}\left[ \sum_{i}p_{i}\left\langle k\right\vert \sqrt{%
\rho _{Ai}}\left\vert k\right\rangle \left\langle k^{\prime }\right\vert 
\sqrt{\rho _{Bi}}\left\vert k^{\prime }\right\rangle \right] ^{2}  \notag \\
&\geqslant &\sum_{kk^{\prime }}\left[ \sum\limits_{i}p_{i}\left\langle
kk^{\prime }\right\vert \left. \psi _{i}\right\rangle _{AB}\left\langle \psi
_{i}\right. \left\vert kk^{\prime }\right\rangle \right] ^{2}
\end{eqnarray}%
with $\rho _{Ai/Bi}$ being the reduced matrix of $\left\vert \psi
_{i}\right\rangle _{AB}\left\langle \psi _{i}\right\vert $ by tracing over A
or B. Based on the Cauchy-Schwarz inequality, we have 
\begin{eqnarray}
&&\sum_{kk^{\prime }}\sum_{i}p_{i}\left\langle k\right\vert \sqrt{\rho _{Ai}}%
\left\vert k\right\rangle ^{2}\sum_{i}p_{i}\left\langle k^{\prime
}\right\vert \sqrt{\rho _{Bi}}\left\vert k^{\prime }\right\rangle ^{2} 
\notag \\
&\geqslant &\sum_{kk^{\prime }}\left\langle kk^{\prime }\right\vert \rho
_{AB}\left\vert kk^{\prime }\right\rangle ^{2}.  \label{bee}
\end{eqnarray}%
Based on the joint concavity of the function $f\left( A,B\right) =TrX^{\dag
}A^{t}XB^{1-t}$ on both $A$ and $B$ (Lieb's theorem) \cite{qip}$,$ Eq. (\ref%
{bee}) becomes
\begin{equation}
\sum_{kk^{\prime }}\left\langle k\right\vert \sqrt{\rho _{A}}\left\vert
k\right\rangle ^{2}\left\langle k^{\prime }\right\vert \sqrt{\rho _{B}}%
\left\vert k^{\prime }\right\rangle ^{2}\geqslant \sum_{kk^{\prime
}}\left\langle kk^{\prime }\right\vert \rho _{AB}\left\vert kk^{\prime
}\right\rangle ^{2}
\end{equation}%
with $\rho _{A/B}$ denoting the reduced matrices of $\rho _{AB}$. So we have%
\begin{eqnarray}
&&[1-C(\rho _{A})][1-C(\rho _{B})]  \notag \\
&\geq &\sum_{kk^{\prime }}\left\langle kk^{\prime }\right\vert \rho
_{AB}\left\vert kk^{\prime }\right\rangle ^{2}=Tr\rho
_{AB}^{2}-C_{l_{2}}(\rho _{AB}),  \label{suup1}
\end{eqnarray}%
where $C_{l_{2}}(\rho _{AB})=Tr\rho _{AB}^{2}-\sum_{kk^{\prime
}}\left\langle kk^{\prime }\right\vert \rho _{AB}\left\vert kk^{\prime
}\right\rangle ^{2}$ is the coherence measure based on the $l_{2}$ norm. One
can easily find that Eq. (\ref{suup1}) will be reduced to theorem 2 if $\rho
_{AB}$ is a pure state. In addition, in order to use the coherence to
describe $\sum_{kk^{\prime }}\left\langle kk^{\prime }\right\vert \rho
_{AB}\left\vert kk^{\prime }\right\rangle ^{2}$ or its lower bound, we now
consider the eigen-decomposition of $\rho _{AB}$, i.e., $\rho
_{AB}=\sum_{i}\lambda _{i}\left\vert \psi _{i}\right\rangle
_{AB}\left\langle \psi _{i}\right\vert $. Thus $\sum_{kk^{\prime
}}\left\langle kk^{\prime }\right\vert \rho _{AB}\left\vert kk^{\prime
}\right\rangle ^{2}$ can be rewritten as%
\begin{gather}
\sum_{kk^{\prime }}\left\langle kk^{\prime }\right\vert \rho _{AB}\left\vert
kk^{\prime }\right\rangle ^{2}=\sum_{kk^{\prime }}\left( \sum_{i}\lambda
_{i}\left\langle kk^{\prime }\right\vert \left. \psi _{i}\right\rangle
_{AB}\left\langle \psi _{i}\right. \left\vert kk^{\prime }\right\rangle
\right) ^{2}  \notag \\
\geq \sum_{kk^{\prime }}\left( \sum_{i}\sqrt{\lambda _{\min }}\sqrt{\lambda
_{i}}\left\langle kk^{\prime }\right\vert \left. \psi _{i}\right\rangle
_{AB}\left\langle \psi _{i}\right. \left\vert kk^{\prime }\right\rangle
\right) ^{2}  \notag \\
=\lambda _{\min }\left( 1-C(\rho _{AB})\right) ,  \label{imp1}
\end{gather}%
where $\lambda _{\min }$ is the minimal nonzero eigenvalue of $\rho _{AB}$.
This is the first conclusion in Corollary 1. It can be seen that Eq. (\ref%
{imp1}) will go back to theorem 2 due to $\lambda _{\min }=1$ for the pure $%
\rho _{AB}$.

Consider the eigen-decomposition of $\rho _{AB}=\sum_{i}\lambda
_{i}\left\vert \psi _{i}\right\rangle _{AB}\left\langle \psi _{i}\right\vert 
$, one can obtain a series of equations akin to Eq. (\ref{iminr}).
Multiplying $\sqrt{\lambda _{i}}$ on both sides of these equations and then
sum over all $i$, we will have 
\begin{eqnarray}
&&\sum_{i}\sqrt{\lambda _{i}}\left\langle k\right\vert \sqrt{\rho _{Ai}}%
\left\vert k\right\rangle \left\langle k^{\prime }\right\vert \sqrt{\rho
_{Bi}}\left\vert k^{\prime }\right\rangle  \notag \\
&\geqslant &\sum\limits_{i}\sqrt{\lambda _{i}}\left\langle kk^{\prime
}\right\vert \left. \psi _{i}\right\rangle _{AB}\left\langle \psi
_{i}\right. \left\vert kk^{\prime }\right\rangle .  \label{haha}
\end{eqnarray}%
Squaring both sides of Eq. (\ref{haha}) and summing over all the $kk^{\prime
}$, we arrive at

\begin{eqnarray}
&&\left( \sum_{i}\sqrt{\lambda _{i}}\left\langle k\right\vert \sqrt{\rho
_{Ai}}\left\vert k\right\rangle \left\langle k^{\prime }\right\vert \sqrt{%
\rho _{Bi}}\left\vert k^{\prime }\right\rangle \right) ^{2}  \notag \\
&\geqslant &\left( \sum\limits_{i}\sqrt{\lambda _{i}}\left\langle kk^{\prime
}\right\vert \left. \psi _{i}\right\rangle _{AB}\left\langle \psi
_{i}\right. \left\vert kk^{\prime }\right\rangle \right) ^{2}.  \label{haha1}
\end{eqnarray}%
According to the Cauchy-Schwarz inequality, Eq. (\ref{haha1}) becomes 
\begin{equation}
\sum_{k}\left\langle k\right\vert \sqrt{\rho _{A}}\left\vert k\right\rangle
^{2}\sum\limits_{k^{\prime }i}\left\langle k^{\prime }\right\vert \sqrt{\rho
_{Bi}}\left\vert k^{\prime }\right\rangle ^{2}\geqslant \sum_{kk^{\prime
}}\left\langle kk^{\prime }\right\vert \sqrt{\rho _{AB}}\left\vert
kk^{\prime }\right\rangle ^{2}  \label{pree1}
\end{equation}%
and 
\begin{equation}
\sum_{ki}\left\langle k\right\vert \sqrt{\rho _{Ai}}\left\vert
k\right\rangle ^{2}\sum\limits_{k^{\prime }}\left\langle k^{\prime
}\right\vert \sqrt{\rho _{B}}\left\vert k^{\prime }\right\rangle
^{2}\geqslant \sum_{kk^{\prime }}\left\langle kk^{\prime }\right\vert \sqrt{%
\rho _{AB}}\left\vert kk^{\prime }\right\rangle ^{2}.  \label{pree2}
\end{equation}%
A simple algebra can further show that Eq. (\ref{pree1}) leads to 
\begin{equation}
\lbrack 1-C(\rho _{A})]\left[ r-\sum\limits_{i}C\left( \rho _{Bi}\right) %
\right] \geqslant 1-C(\rho _{AB})  \label{eq1}
\end{equation}%
and Eq. (\ref{pree2}) leads to%
\begin{equation}
\lbrack r-\sum\limits_{i}C(\rho _{Ai})]\left[ 1-C\left( \rho _{B}\right) %
\right] \geqslant 1-C(\rho _{AB}).  \label{eq2}
\end{equation}%
Combine Eqs. (\ref{eq1}) and (\ref{eq2}), one will obtain a symmetric form%
\begin{equation}
\lbrack 1-C(\rho _{A})][1-C(\rho _{B})]\geq \frac{1}{c_{s}}\left[ 1-C(\rho
_{AB})\right] ^{2},
\end{equation}%
where $r$ denotes the rank of $\rho _{AB}$ and $c_{s}=[r-\sum\limits_{i}C(%
\rho _{Ai})][r-\sum\limits_{i}C(\rho _{Bi})]$ with $\sum\limits_{i}C(\rho
_{Ai/Bi})$ corresponding to the sum of the subsystematic (A/B) coherence of
all the eigenstates. It is obvious that the inequality will be reduced to
the case of pure states for pure $\rho _{AB}$. The proof of Corollary 1 is
finished.

\subsection{Proof of Corollary 2}

Corollary 2 is the result of the direct application of Corollary 1, so it is
sufficient to consider an example to demonstrate how to arrive at the
expected inequalities and how to determine the coefficient $\lambda _{M}$
and $c_{sT}$. Without loss of generality, let's consider a quardripartite
quantum state $\rho _{ABCD}$. At first, we would like to consider $\rho
_{ABCD}$ as a bipartite state as $\rho _{\left( AB\right) \left( CD\right) }$
(or $\rho _{A\left( BCD\right) }$ and so on). Based on Corollary 1, we have 
\begin{equation}
\left[ 1-C(\rho _{AB})\right] \left[ 1-C(\rho _{CD})\right] \geq \lambda
_{\min 1}\left[ 1-C\left( \rho _{ABCD}\right) \right] ,  \label{c21}
\end{equation}%
where $\rho _{AB}$ and $\rho _{CD}$ are the reduced density matrices of $%
\rho _{\left( AB\right) \left( CD\right) }$ and $\lambda _{\min 1}$ is the
minimal nonzero eigenvalue of $\rho _{\left( AB\right) \left( CD\right) }$.
One can also find the similar results for $\rho _{AB}$ and $\rho _{CD}$,
that is,
\begin{eqnarray}
\left[ 1-C(\rho _{A})\right] \left[ 1-C(\rho _{B})\right] &\geq &\lambda
_{\min 2}\left[ 1-C(\rho _{AB})\right] ,  \label{c22} \\
\left[ 1-C(\rho _{C})\right] \left[ 1-C(\rho _{D})\right] &\geq &\lambda
_{\min 3}\left[ 1-C(\rho _{CD})\right] ,  \label{c23}
\end{eqnarray}%
where $\lambda _{\min 2}$ and $\lambda _{\min 3}$ are the minimial nonzero
eigenvalues for $\rho _{AB}$ and $\rho _{CD}$, respectively. Thus one can
stop at Eq. (\ref{c21}) where $\lambda _{M}=\lambda _{\min 1}$. One can
combine Eq. (\ref{c21}) and Eq. (\ref{c22}) and obtain%
\begin{eqnarray}
&&\left[ 1-C(\rho _{A})\right] \left[ 1-C(\rho _{B})\right] \left[ 1-C(\rho
_{CD})\right]  \notag \\
&\geq &\lambda _{\min 1}\lambda _{\min 2}\left[ 1-C\left( \rho
_{ABCD}\right) \right] ,
\end{eqnarray}%
where $\lambda _{M}=\lambda _{\min 1}\lambda _{\min 2}$. The similar
conclusion can be got if Eq. (\ref{c21}) and Eq. (\ref{c23}) are combined.
Of course, one can combine all the three equations, and finally get to 
\begin{equation}
\prod\limits_{i=A,B,C,D}\left[ 1-C(\rho _{i})\right] \geq \lambda _{M}\left[
1-C\left( \rho _{ABCD}\right) \right]  \label{ctt}
\end{equation}%
with $\lambda _{M}=\lambda _{\min 1}\lambda _{\min 2}\lambda _{\min 3}.$This
demonstrates how to obtain Eq. (24) in the main text.

Let's consider $\rho _{ABCD}$ again and first look at it as a bipartite
state, for example, $\rho _{A\left( BCD\right) }$. Based on Corollary 1, we
have 
\begin{equation}
\left[ 1-C(\rho _{A})\right] \left[ 1-C(\rho _{BCD})\right] \geq \frac{1}{%
c_{s1}}\left[ 1-C\left( \rho _{ABCD}\right) \right] ^{2},  \label{tt1}
\end{equation}%
where $c_{s1}=\left[ r_{1}-\sum\limits_{i=1}^{r_{1}}C(\rho _{Ai})\right] %
\left[ r_{1}-\sum\limits_{i=1}^{r_{1}}C(\rho _{(BCD)i}\right] $ with $\rho
_{Ai}$ and $\rho _{(BCD)i}$ denoting the reduced density matrices of \textit{%
i}th eigenstate of $\rho _{ABCD}$ and $r_{1}$ being the rank of $\rho
_{ABCD} $. If one just wants to consider such a bipartite grouping, Eq. (\ref%
{tt1}) is the final description of polygamy with $c_{sT}=c_{s1}$ and $%
n_{1}=n_{2}=1. $ One can continue to consider $\rho _{BCD}$ as a bipartite
state $\rho _{\left( BC\right) D}$ and continue to use Corollary 1. Then we
will obtain 
\begin{equation}
\left[ 1-C(\rho _{BC})\right] \left[ 1-C(\rho _{D})\right] \geq \frac{1}{%
c_{s2}}\left[ 1-C\left( \rho _{BCD}\right) \right] ^{2},  \label{tt2}
\end{equation}%
where $c_{s2}=\left[ r_{2}-\sum\limits_{i=1}^{r_{2}}C(\rho _{(BC)i})\right] %
\left[ r_{2}-\sum\limits_{i=1}^{r_{2}}C(\rho _{Di})\right] $ with $\rho
_{(BC)i}$ and $\rho _{Di}$ representing the reduced density matrices of 
\textit{i}th eigenstate of $\rho _{BCD}$ and $r_{2}$ being the rank of $\rho
_{BCD}$. Substitute Eq. (\ref{tt2}) into Eq. (\ref{tt1}), one will arrive at%
\begin{eqnarray}
&&\left[ 1-C(\rho _{A})\right] \sqrt{\left[ 1-C(\rho _{BC})\right] \left[
1-C(\rho _{D})\right] }  \notag \\
&\geq &\frac{1}{c_{s1}c_{s2}}\left[ 1-C\left( \rho _{ABCD}\right) \right]
^{2},  \label{tt3}
\end{eqnarray}%
with $c_{sT}=c_{s1}c_{s2}$. Thus we can see that $n_{1}=1,n_{2}=n_{3}=\frac{1%
}{2}$. Of course, one can continute to divide $\rho _{BC}$ and obtain
another inequality, which is omitted here.
\begin{figure}[tbp]
\centering
\includegraphics[width=1\columnwidth]{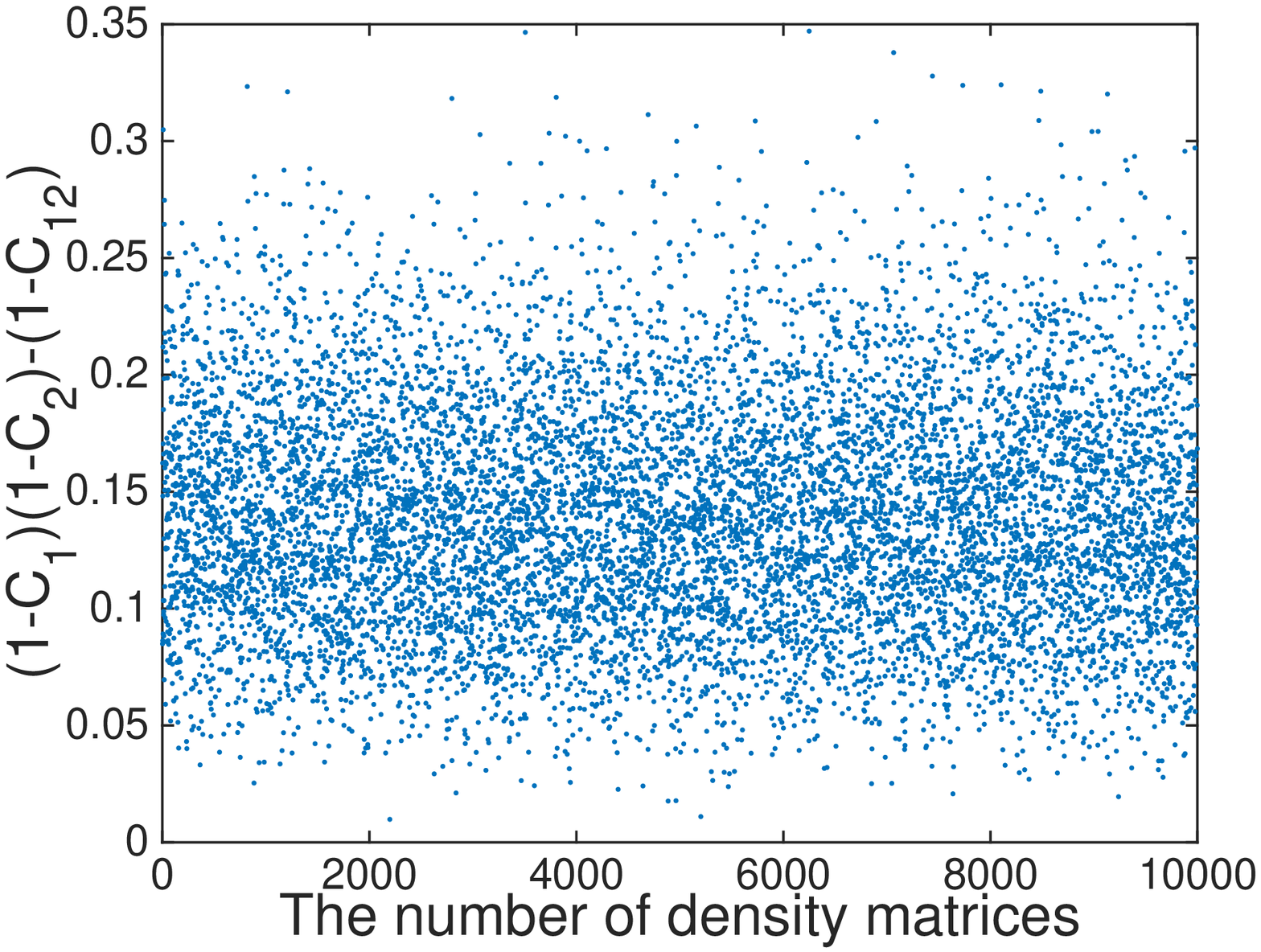} \centering
\caption{All the density matrices $\protect\rho _{AB}$ are generated in $%
(2\otimes 3)$-dimensional Hilbert space. }
\end{figure}
\begin{figure}[tbp]
\centering
\includegraphics[width=1\columnwidth]{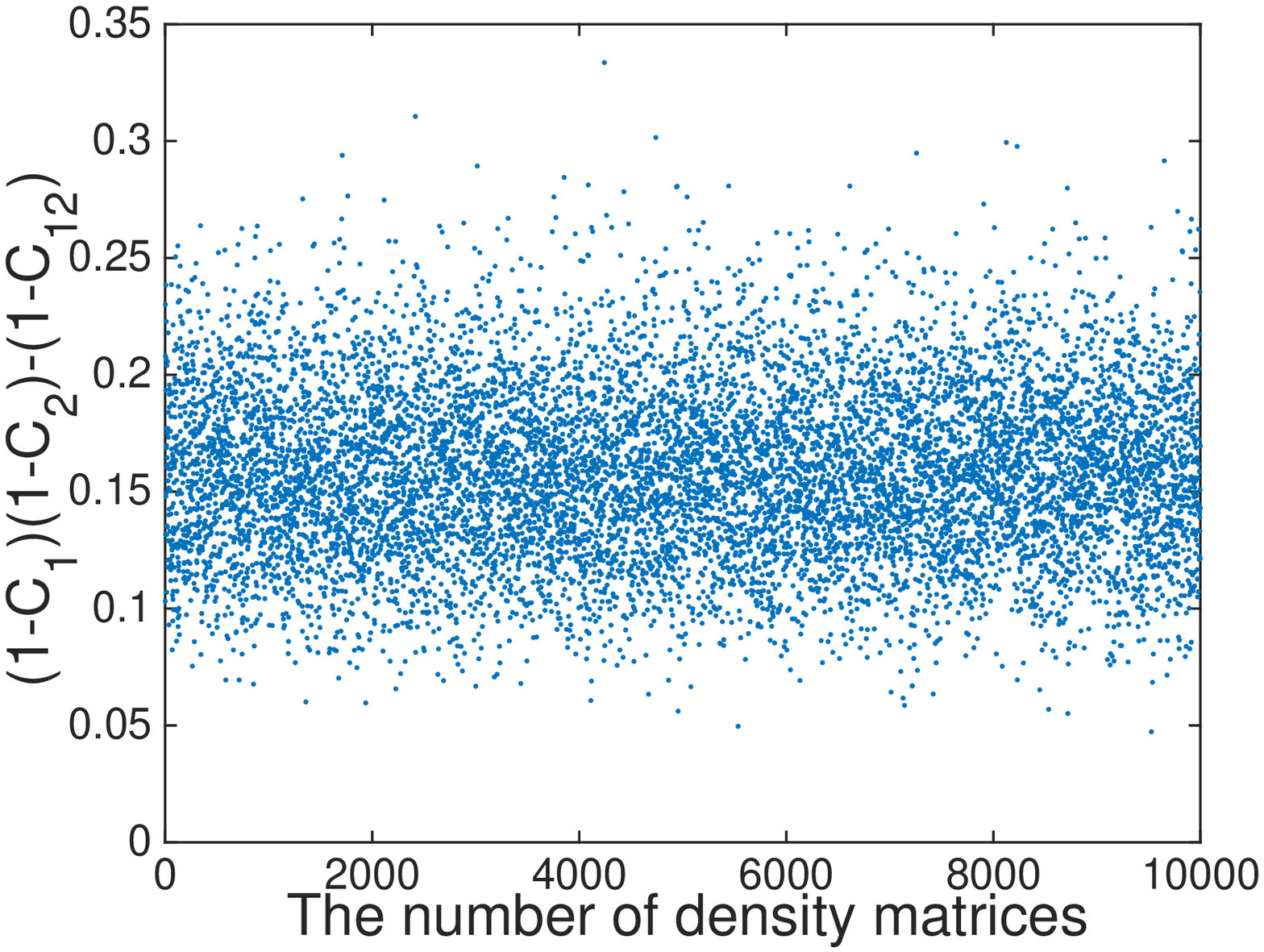} \centering
\caption{All the density matrices $\protect\rho _{AB}$ are generated in $%
(3\otimes 3)$-dimensional Hilbert space.}
\end{figure}
\begin{figure}[tbp]
\centering
\includegraphics[width=1\columnwidth]{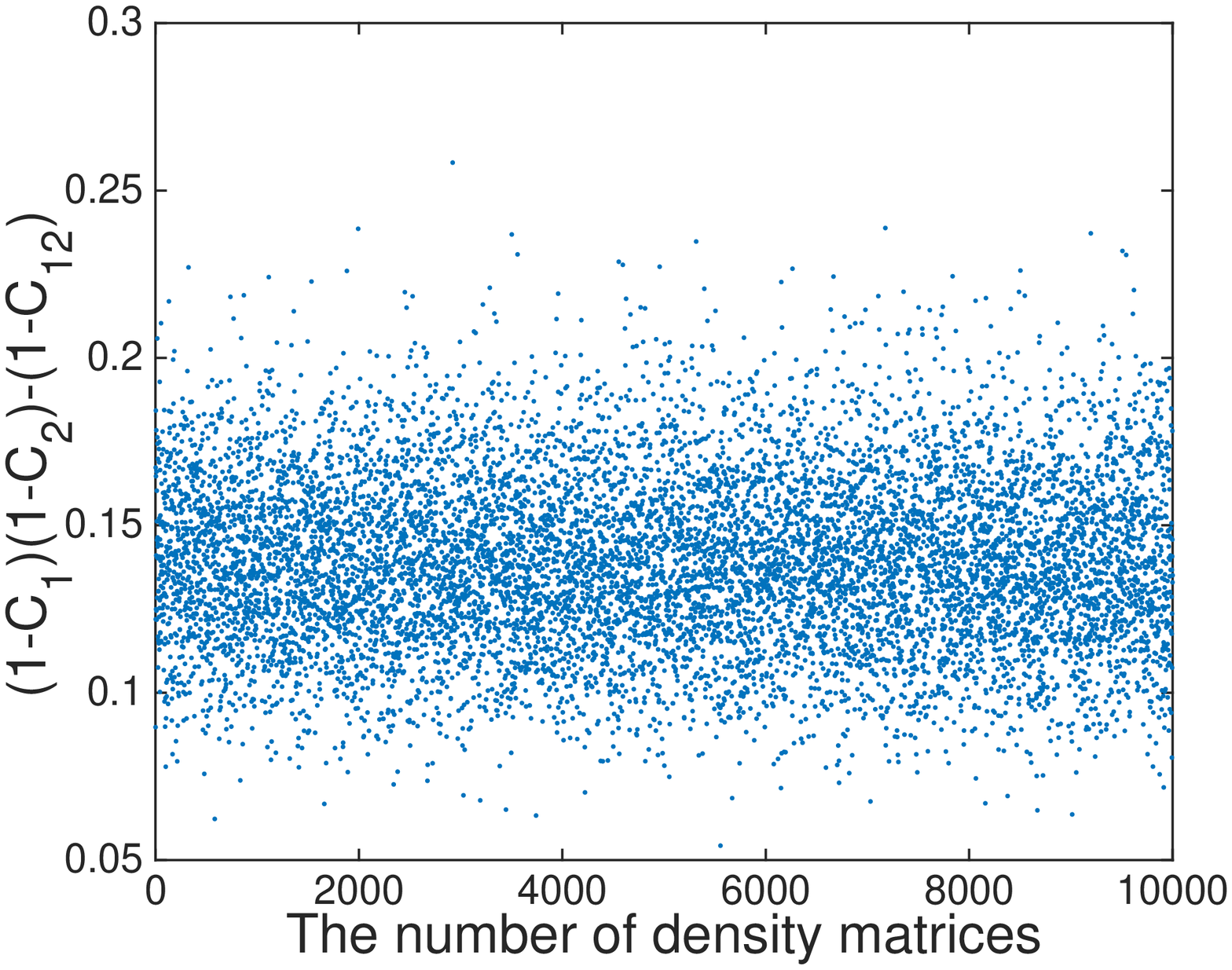} \centering
\caption{All the density matrices $\protect\rho _{AB}$ are generated in $%
(3\otimes 4)$-dimensional Hilbert space. }
\end{figure}
\begin{figure}[tbp]
\centering
\includegraphics[width=1\columnwidth]{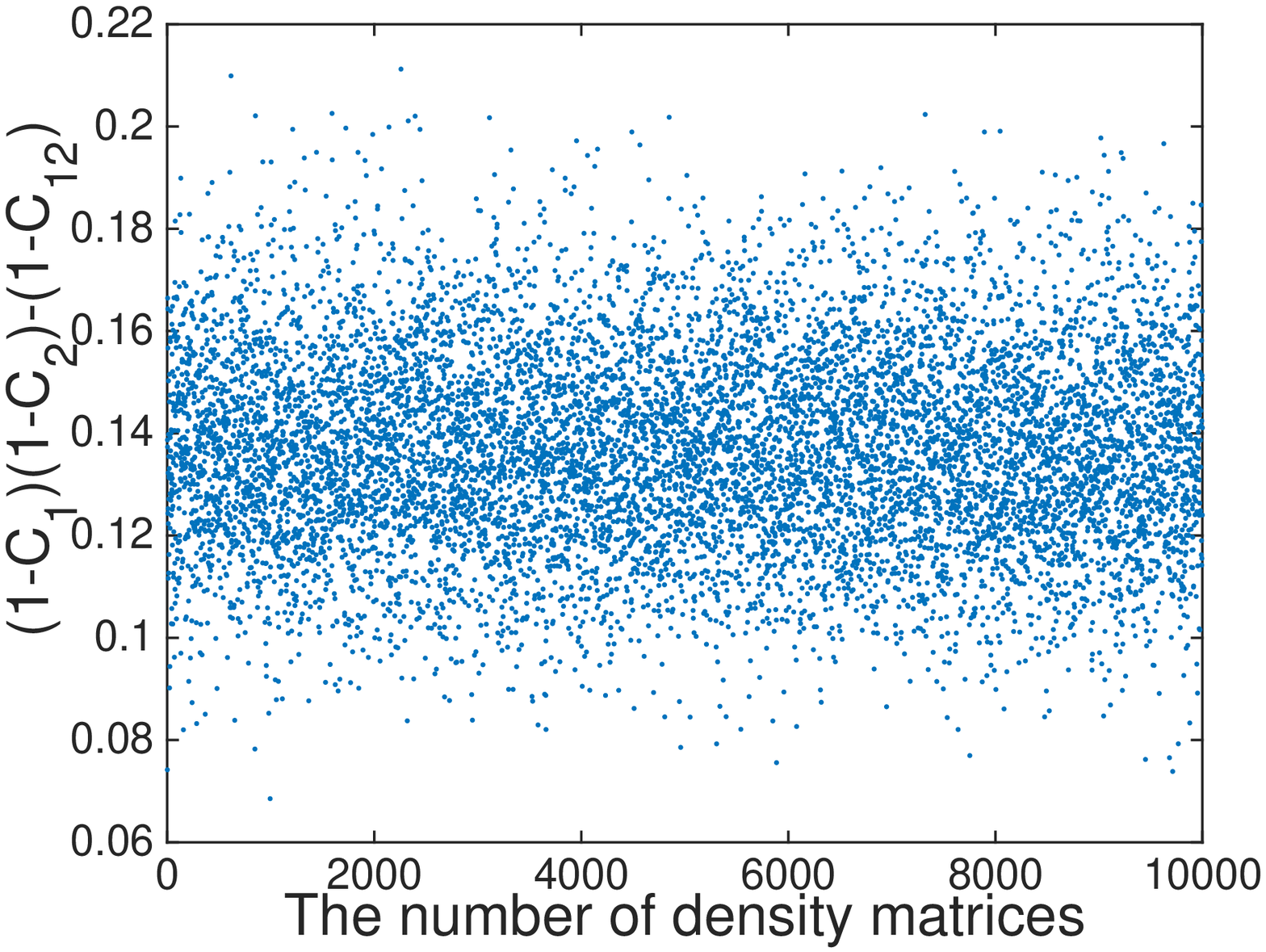} \centering
\caption{All the density matrices $\protect\rho _{AB}$ are generated in $%
(4\otimes 4)$-dimensional Hilbert space. }
\end{figure}

\section{The measurable relative-entropy coherence }\label{app:measure}

Now we  show that the relative-entropy coherence $C_r\left(\rho\right)$ can
be directly measured in experiment.

$C_{r}\left( \rho \right) $ can be written as 
\begin{eqnarray}
C_{r}\left( \rho \right)  &=&S\left( \rho ^{\star }\right) -S\left( \rho
\right)   \notag \\
&=&\sum_{j}\lambda _{j}\log {\lambda _{j}}-\sum_{k}\rho _{kk}\log {\rho _{kk}%
}
\end{eqnarray}%
where $\rho ^{\star }$ denotes the state by deleting all off-diagonal
entries of $\rho $, the $\lambda _{j}$'s represent the eigenvalues of $\rho $
and $\rho _{kk}=\left\langle k\right\vert \rho \left\vert k\right\rangle $
is the diagonal entries of $\rho $ within the reference basis $\{\left\vert
k\right\rangle \}$ . It is obvious that once the knowledge on $\lambda _{j}$
and $\rho _{kk}$ are extracted from an experiment, $C\left( \rho \right) $
is determined. This can be accomplished by the generalized standard overlap
measurement \cite{en4,overlap} and simple projective measurements. To do so,
we can define the generalized swapping operator $V_{n}$ for natural number $%
n>1$ as $V_{n}\left\vert \psi _{1},\psi _{2},\cdot \cdot \cdot ,\psi
_{n}\right\rangle =\left\vert \psi _{n},\psi _{1},\psi _{2},\cdot \cdot
\cdot ,\psi _{n-1}\right\rangle $. So a controlled $V_{n}$ gate can be
constructed as $\mathbb{I}_{2}\oplus V_{n}$ with a qubit as the control
qubit. It is easy to find that $Tr\rho ^{n}=TrV_{n}\rho ^{\otimes n}$. Now
let's first prepare a probing qubit $\left\vert \varphi \right\rangle _{p}=%
\frac{1}{\sqrt{2}}\left( \left\vert 0\right\rangle +\left\vert
1\right\rangle \right) $ and $n$ copies of measured state $\rho $. Then let
the $n+1$ particles undergo the controlled $V_{n}$ gate. Finally, let's
measure $\sigma _{x}$ on the probing qubit and obtain $\pm 1$ with the
probability $p_{n}^{\pm }=\frac{1\pm Tr\rho ^{n}}{2}$. Thus based on $%
p_{n}^{+}$ (or $p_{n}^{-}$) for $n=2,3,\cdots ,N_{D}$, with $Tr\rho =1$ all
the $\lambda _{j}$'s can be unambiguously determined and so are $\sqrt{%
\lambda _{j}}$'s. In addition, $\left\langle k\right\vert \rho \left\vert
k\right\rangle $ can be measured directly by the projective measurement
subject to the projectors $\hat{P}_{k}=\left\vert k\right\rangle
\left\langle k\right\vert $. Therefore, $C\left( \rho \right) $ is obtained.
Compared with $N_{D}^{2}-1$ observables in QST, the total cost is $N_{D}-1$
controlled $V_{n}$ gates plus $N_{D}-1$ projective measurements assisted by
at most $N_{D}$ copies of the state.

\section{The conjecture}\label{app:conj}
The polygamy relation has an elegant form for the bipartite pure state, but
one can easily find that such a relation doesn't hold for general mixed
states. This can be seen as follows. Let's consider the qubit state $%
\rho_{AB}=p\left\vert
\psi_1\right\rangle\left\langle\psi_1\right\vert+(1-p)\left\vert\psi_2\right%
\rangle\left\langle\psi_2\right\vert$  with $\left\vert\psi_1\right%
\rangle=[-0.5612,-0.982,0.8119,0.1272]^T$,  $\left\vert\psi_2\right%
\rangle=[0.8006,0.1842,0.5556,0.1283]^T$ and $\left\langle\psi_1\right.\left%
\vert\psi_2\right\rangle=0$, $p=0.0443.$ A simple algebra can show that $%
C(\rho_1)=0.2582$, $C(\rho_2)=0.0909$ and $C(\rho)=0.3242$ with $\rho_i=%
\mathrm{Tr}_{A/B}\left\vert\psi_i\right\rangle\left\langle\psi_i\right\vert$%
. Thus it is easy to check that $(1-C(\rho_1))(1-C(\rho_2))=0.7418\times
0.9091=0.6744<0.6758=1-C(\rho)$.  However, through our numerical test, we
conjecture that the same form of our theorem 2 for $\left( N\geq 6\right) $%
-dimensional state could also be satisfied. In Fig. 1, Fig. 2, Fig. 3 and
Fig. 4, we numerically test the inequality in high dimensional systems, but
we don't find the counter-example. In the figures, we use $C_{12}$ to denote
the bipartite state $C(\rho_{AB})$ and $C_i$ to denote $C(\rho_i)$ with $%
\rho_i=\mathrm{Tr}_{A/B}\rho_{AB}$ representing the corresponding reduced
density matrices. All the tested density matrices $\rho _{AB}=\frac{(A\ast
A^{\prime }+B\ast B^{\prime })}{\mathrm{Tr}A\ast A^{\prime }+B\ast B^{\prime
}}$ with $B=C+iD$ and $A,C,D$ randomly generated by Matlab R2014b. One can
find that in all the figures $(1-C_1)(1-C_2)-(1-C_{12})\geq 0$. Comparing
the four figures, one can find that the minimal value of $%
(1-C_1)(1-C_2)-(1-C_{12})$ in the figures is increased with the increasing
of the dimension of the state. In this sense, we would like to conjecture
that this relation should be satisfied in $(N\geq 6)$-dimensional systems.




\begin{thebibliography}{99}
\bibitem{horoe} R. Horodecki, P. Horodecki, M. Horodecki, and K. Horodecki,
Rev. Mod. Phys. \textbf{81}, 865 (2009).

\bibitem{disc1} H. Ollivier, and W. H. Zurek, Phys. Rev. Lett. \textbf{88},
017901 (2001).

\bibitem{disc2} L. Henderson, and V. Vedral, J. Phys. A.: Math. Gen. \textbf{%
34}, 6899 (2001).

\bibitem{bell} J. S. Bell, Rev. Mod. Phys. \textbf{38}, 447 (1966).

\bibitem{Engel} G. S. Engel, T. R. Calhoun, E. L. Read, T.-K. Ahn, T. Man%
\v{c}al, Y.-C. Cheng, R. E. Blankenship, and G. R. Fleming, Nature (London) 
\textbf{446}, 782 (2007).

\bibitem{Plenio} M. B. Plenio, and S. F. Huelga, New J. Phys. \textbf{10},
113019 (2008).

\bibitem{Coll} E. Collini, C. Y. Wong, K. E. Wilk, P. M. G. Curmi, P.
Brumer, and G.D. Scholes, Nature (London) \textbf{463}, 644 (2010).

\bibitem{loyd} S. Lloyd, J. Phys. Conf. Ser. \textbf{302}, 012037 (2011).

\bibitem{licm} C. M. Li, N. Lambert, Y.-N. Chen, G. Y. Chen, and F. Nori,
Sci. Rep. \textbf{2}, 885 (2012).

\bibitem{Huel} S. Huelga, and M. Plenio, Contemp. Phys. \textbf{54}, 181
(2013).

\bibitem{Ryb} L. Rybak, S. Amaran, L. Levin, M. Tomza, R. Moszynski, R.
Kosloff, C. P. Koch, and Z. Amitay, Phys. Rev. Lett. \textbf{107}, 273001
(2011).

\bibitem{berg} J. \AA berg, Phys. Rev. Lett. \textbf{113}, 150402 (2014).

\bibitem{Nar} V. Narasimhachar, and G. Gour, arXiv: 1409.7740 [quant-ph].

\bibitem{Horo} P. \'{C}wikli\'{n}ski, M. Studzi\'{n}ski, M. Horodecki, and
J. Oppenheim, arXiv: 1405.5029 [quant-ph].

\bibitem{Los1} M. Lostaglio, D. Jennings, and T. Rudolph, Nat. Commun. 
\textbf{6}, 6383 (2015).

\bibitem{Los2} M. Lostaglio, K. Korzekwa, D. Jennings, and T. Rudolph, Phys.
Rev. X \textbf{5}, 021001 (2015).

\bibitem{reb} P. Rebentrost, M. Mohseni, and A. Aspuru-Guzik, J. Phys. Chem.
B \textbf{113}, 9942 (2009).

\bibitem{wit} B. Witt, and F. Mintert, New J. Phys. 15, 093020 (2013).

\bibitem{Vaz} H. Vazquez, R. Skouta, S. Schneebeli, M. Kamenetska, R.
Breslow, L. Venkataraman, and M. Hybertsen, Nat. Nano-technol. \textbf{7},
663 (2012).

\bibitem{Karl} O. Karlstr\"{o}m, H. Linke, G. Karlstr\"{o}m, and A. Wacker,
Phys. Rev. B \textbf{84}, 113415 (2011).

\bibitem{Pleniom} T. Baumgratz, M. Cramer, and M. B. Plenio, Phys. Rev.
Lett. \textbf{113}, 140401 (2014).

\bibitem{Lewen} S. Rana, P. Parashar, and M. Lewenstein, Phys. Rev. A 
\textbf{93}, 012110 (2016).

\bibitem{Giro} D. Girolami, Phys. Rev. Lett. \textbf{113}, 170401 (2014).

\bibitem{Napoli} C. Napoli, T. R. Bromley, M. Cianciaruso, M. Piani, N.
Johnston, and G. Adesso, Phys. Rev. Lett. \textbf{116}, 150502 (2016).

\bibitem{Yu09} C. S. Yu, and H. S. Song, Phys. Rev. A \textbf{80}, 022324
(2009).

\bibitem{Ma} J. Ma, B. Yadin, D. Girolami, V. Vedral, and M. Gu, Phys. Rev.
Lett. \textbf{116}, 160407 (2016).

\bibitem{Stre} A. Streltsov, U. Singh, H. S. Dhar, M. N. Bera, and G.
Adesso, Phys. Rev. Lett. \textbf{115}, 020403 (2015).

\bibitem{yuto} C. S. Yu, S. R. Yang, and B. Q. Guo, Quant. Inf. Proc. 
\textbf{15}, 3773 (2016).


\bibitem{Winter} A. Winter, and D. Yang, Phys. Rev. Lett. \textbf{116},
120404 (2016).

\bibitem{Chi} E. Chitambar, A. Streltsov, S. Rana, M. N. Bera, G. Adesso,
and M. Lewenstein, Phys. Rev. Lett. \textbf{116}, 070402 (2016).

\bibitem{Chi2} E. Chitambar, and M.-H. Hsieh, Phys. Rev. Lett. \textbf{117},
020402 (2016).

\bibitem{Chi3} E. Chitambar, and Gilad Gour, Phys. Rev. Lett. \textbf{117},
030401 (2016).

\bibitem{Marvian} I. Marvian, and R. W. Spekkens, Phys. Rev. A \textbf{90},
062110 (2014).

\bibitem{Marvian2} I. Marvian, R. W. Spekkens, and P. Zanardi, Phys. Rev. A 
\textbf{93}, 052331 (2016).

\bibitem{Yao} Y. Yao, X. Xiao, L. Ge, and C. P. Sun, Phys. Rev. A \textbf{92}%
, 022112 (2015).

\bibitem{Sing} U. Singh, L. Zhang, and A. K. Pati, Phys. Rev. A \textbf{93},
032125 (2016).

\bibitem{Rast} A. E. Rastegin, Phys. Rev. A \textbf{93}, 032136 (2016).

\bibitem{Piani} M. Piani, M. Cianciaruso, T. R. Bromley, C. Napoli, N.
Johnston, and G. Adesso, Phys. Rev. A \textbf{93}, 042107 (2016).

\bibitem{Radha} C. Radhakrishnan, M. Parthasarathy, S. Jambulingam, and T.
Byrnes, Phys. Rev. Lett. \textbf{116}, 150504 (2016).


\bibitem{tom1} A. G. White, D. F. V. James, P. H. Eberhard, and P. G. Kwiat,
Phys. Rev. Lett. \textbf{83}, 3103 (1999).

\bibitem{tom2} H. H\"{a}ffner et al., Nature (London) \textbf{438}, 643
(2005).


\bibitem{skew1} E. P. Wiger, and M. M. Yanase, Proc. Natl. Acad. Sci. 49,
910 (1963).

\bibitem{skew2} E. H. Lieb, Adv. Math. \textbf{11}, 267 (1973).

\bibitem{skewl} S. Luo, Phys. Rev. Lett. \textbf{91}, 180403 (2003).

\bibitem{fisher1} C. W. Helstrom, \textit{Quantum detection and estimation
theory} ( Academic Press, New York, 1976).

\bibitem{fisher2} A. S. Holevo, \textit{Probabilistic and statistical
aspects of quantum theory} (North-Holland, Am- sterdam, 1982).

\bibitem{mono1} V. Coffman, J. Kundu, and W. K. Wootters, Phys. Rev. A 
\textbf{61}, 052306 (2000).

\bibitem{mono2} T. J. Osborne, and F. Verstraete, Phys. Rev. Lett. \textbf{96%
}, 220503 (2006).

\bibitem{mono3} C. S. Yu, and H. S. Song, Phys. Rev. A \textbf{77}, 032329
(2008).


\bibitem{Mal} L. Mandel, and E. Wolf, \textit{Optical Coherence and Quantum
Optics} (Cambridge University Press, Cambridge, England, 1995).

\bibitem{luo} S. Luo, and Q. Zhang, Phys. Rev. A \textbf{69}, 032106 (2004).


\bibitem{fis2} U. Dorner, R. Demkowicz-Dobrzanski, B. J. Smith, J. S.
Lundeen, W.Wasilewski, K. Banaszek, and I. A. Walmsley, Phys. Rev. Lett. 
\textbf{102}, 040403 (2009).

\bibitem{fis1} S. L. Braunstein and C. M. Caves, Phys. Rev. Lett. \textbf{72}%
, 3439 (1994);

\bibitem{fis11} S. L. Braunstein, C. M. Caves, and G. J. Milburn, Ann. Phys.
(N.Y.)\textbf{247}, 135 (1996).

\bibitem{luo1} S. L. Luo, Proc. Am. Math. Soc. \textbf{132}, 885 (2003).

\bibitem{review} K. Modi, A. Brodutch, H. Cable, T. Paterek, and V. Vedral,
Rev. Mod. Phys. \textbf{84}, 1655 (2012).

\bibitem{lqu} D. Girolami, T. Tufarelli, and G. Adesso, Phys. Rev. Lett. 
\textbf{110}, 240402 (2013).

\bibitem{Yuquan} C. S. Yu, S. Wu, X. G. Wang, X. X. Yi, and H. S. Song,
Europhys. Lett. \textbf{107}, 10007 (2014).

\bibitem{en1} F. Mintert, M. Ku\`{s}, and A. Buchleitner, Phys. Rev. Lett. 
\textbf{95}, 260502 (2005).

\bibitem{en2} C. S. Yu, and H. S. Song, Phys. Rev. A \textbf{76}, 022324
(2007).

\bibitem{en3} J. M. Cai, and W. Song, Phys. Rev. Lett. \textbf{101}, 190503
(2008).

\bibitem{dis1} J. S. Jin, F. Y. Zhang, C. S. Yu, and H. S. Song, J. Phys. A:
Math. Theor. \textbf{45}, 115308 (2012).

\bibitem{dis2} D. Girolami, and G. Adesso, Phys. Rev. Lett. \textbf{108},
150403 (2012).

\bibitem{en4} T . A. Brun, Quant. Inf. Comput. \textbf{4}, 401 (2004).

\bibitem{overlap} R. Filip, Phys. Rev. A \textbf{65}, 062320 (2002).

\bibitem{qip} M. A. Nielsen and I. L. Chuang, \textit{Quantum Computation
and Quantum Information} (Cambridge University Press, Cambridge, England,
2010).
\end{thebibliography}
\end{document}